\documentclass[a4paper,usenatbib]{mnras}

\usepackage[T1]{fontenc}
\usepackage{ae,aecompl}

\usepackage{graphicx}	
\usepackage{amsmath}	
\usepackage{amssymb}	

\title[Structural and Dynamical Properties of Galaxies]{Structural and
  Dynamical Properties of Galaxies in a Hierarchical Universe: Sizes
  and Specific Angular Momenta}

\author[A. Zoldan et al.]{
Anna Zoldan,$^{1}$\thanks{E-mail: anna.zoldan@inaf.it}
Gabriella De Lucia,$^{1}$
Lizhi Xie,$^{1}$
Fabio Fontanot$^{1}$ and 
\newauthor
Michaela Hirschmann$^{2}$
\\
$^{1}$OATS, INAF, Via Bazzoni 2, 34124-Trieste, TS, Italy\\
$^{2}$ Institut d'Astrophysique de Paris, Sorbonne Universit\'{e}s, UPM-CNRS, UMR7095, F-75014, Paris, France
}

\date{Accepted XXX. Received YYY; in original form ZZZ}

\pubyear{2017}

\begin{document}
\label{firstpage}
\pagerange{\pageref{firstpage}--\pageref{lastpage}}
\maketitle

\begin{abstract}
We use a state-of-the-art semi-analytic model to study the size and the
specific angular momentum of galaxies. Our model includes a specific treatment
for the angular momentum exchange between different galactic components. Disk
scale radii are estimated from the angular momentum of the gaseous/stellar
disk, while bulge sizes are estimated assuming energy conservation. The
predicted size--mass and angular momentum--mass relations are in fair agreement
with observational measurements in the local Universe, provided a treatment for
gas dissipation during major mergers is included. Our treatment for disk
instability leads to unrealistically small radii of bulges formed through this channel, and predicts an offset between the size--mass relations of
central and satellite early-type galaxies, that is not observed. The model
reproduces the observed dependence of the size--mass relation on morphology, and
predicts a strong correlation between specific angular momentum and cold gas
content. This correlation is a natural consequence of galaxy evolution:
gas-rich galaxies reside in smaller halos, and form stars gradually until
present day, while gas-poor ones reside in massive halos, that formed most of their stars at early epochs, when the angular momentum of their parent halos is
low. The dynamical and structural properties of galaxies can be strongly
affected by a different treatment for stellar feedback, as this would modify
their star formation history. A higher angular momentum for gas accreted
through rapid mode does not affect significantly the properties of massive
galaxies today, but has a more important effect on low-mass galaxies at
higher redshift.
\end{abstract}




\section{Introduction}
\label{sec:intro}
The history of a galaxy (and of its components) is determined by a
network of physical processes that drive a complex exchange of mass,
energy, metals and angular momentum.  In a hierarchical Universe,
galaxies are believed to form from the collapse of baryons in the
potential well of Dark Matter (DM) halos. These acquire their angular
momentum through gravitational tidal torques during the growth of
perturbations 
\citep{peebles1969,white1984,barnes1987}.
Numerical simulations show that the gas and dark matter
within virialized systems have very similar initial angular momentum
distributions \citep[in agreement with standard assumptions,][]{fall1980intro}, although there is a slight misalignment
between their angular momentum vectors
\citep{vandenbosch2002ang_mom,vandenbosch2003ang_mom,
  sharma2005ang_mom}.

As hot gas cools, preserving its angular momentum, a rotationally
supported cold gas disk forms. Star formation takes place within
overdense regions of this cold gas disk, originating a stellar disk
that is assumed to inherit the angular momentum of the cold gas from
which it forms. A spheroidal component (a bulge) can form through
galaxy--galaxy mergers \citep{barnes1991,katz1992, hopkins2010}, or
through dynamical instabilities in disk galaxies \citep[see][and references therein]{kormendy2004PB}.
The former channel is
believed to originate classical and kinetically hot spheroids, while
the latter is believed to lead to the formation of the so-called
`pseudo-bulges' that are dynamically cold and have S\'ersic indices,
stellar populations and velocity dispersions intermediate between
those of classical bulges and disks
\citep{andredakis1994,dejong1996PB,peletier1996PB,carollo2001,macarthur2003PB,kormendy2004PB,drory2007PB}. Therefore,
the observed structure and dynamical properties of galaxies are
intimately connected to the evolution of the gas angular momentum.

Progress on the observational side has recently allowed a systematic
study of the sizes and angular momenta for statistical samples of
galaxies.  High resolution imaging in several photometric bands became
available from e.g. the Sloan Digital Sky Survey (SDSS,
\citealt{york2000sdss}) or the Galaxy And Mass Assembly (GAMA,
\citealt{driver2011gama}) project in the local Universe, and e.g.
CANDELS \citep{grogin2011CANDELS} at higher redshift.  The advent of
Integral Field Spectroscopy allowed measurements of spatially resolved
properties for thousands of galaxies, starting with the pioneering
work done with SAURON \citep{bacon2001sauron} and culminating in ongoing
projects like CALIFA, SAMI, and MaNGA
\citep{sanchez2012califa,bryant2015sami,bundy2015manga}.

The first analysis of the size--mass relation based on a statistical
sample of galaxies was carried out by \citet{shen2003_sdss}, using
SDSS data.  The observed relation exhibits a large scatter that
depends on the morphology of the galaxies, with Late Type (LT)
galaxies having, on average, larger characteristic radii than Early
Types (ETs) at fixed stellar mass.  Different ET/LT selections lead to similar size--mass
median relations \citep[e.g.][based on GAMA]{lange2015}.  Satellite LT
galaxies are slightly smaller than centrals, while the size of ET
galaxies does not appear to be significantly affected by the
environment \citep[e.g.][both based on
  SDSS]{weinmann2009sdss,huertas2013size}.  The sizes of both LT and
ET galaxies tend to decrease with increasing redshift
\citep{ichikawa2012,van_der_wel2014candles}.

The first attempt to study the relation between the specific angular momentum
($j_*$) and galaxy stellar mass ($M_*$) was made by \citet{fall1983j}, for a
small sample of galaxies of different morphological types.
\citet{romanowsky2012js} extended the original analysis using different
techniques  (including long slit spectroscopy, integrated stellar light
  spectroscopy, and extended PN kinematics) to extrapolate the rotational
  profile of LT and ET galaxies out to large aperture radii (from $\sim2$ to
  $\sim8\,R_{eff}$).  They used these results to formulate an empirical
  relation to infer the total specific angular momentum from measurements
  limited to $\sim 2$ half-light radii, and applied this empirical formula to a
  larger sample of galaxies.  This study showed that, as for the sizes, the
scatter in the $j_*$--$M_*$ relation depends on galaxy morphology. Recently,
spatially resolved velocity maps have become available for hundreds of
galaxies. Data from the ATLAS$^{\rm 3D}$ project \citep{cappellari2011atlas3d}
revealed that ET galaxies exhibit a complex internal dynamical structure. ET
galaxies are classified as `fast' or `slow' rotators according to the
importance of the rotational velocity compared to the velocity dispersion,
respectively. This classification has been connected to different galaxy
assembly histories: fast rotators originate predominantly from secular
evolution, while slow rotators are associated with recent and numerous
accretion events
\citep{davis2011atlas3d_mol_gas,serra2014atlas3d_HI,davis2016FR_SR}.
\citet{obreschkow2014jmbt} used a small sample of gas-rich spiral galaxies from
THINGS \citep{walter2008things}, and showed that they lie on a plane in the 3D
space described by the specific angular momentum, the stellar mass, and the
bulge over total mass ratio. An estimate of the $j_*$--$M_*$ relation for both
LT and ET galaxies has been recently provided by \citet{cortese2016sami}, based
on SAMI data \citep{bryant2015sami}. In the near future, it will be possible to
extend these studies to even larger mass selected samples of galaxies using
e.g. data from MaNGA \citep{bundy2015manga}.

Many theoretical studies have focused on the origin of both galaxy sizes and
their specific angular momenta in a cosmological context.  The analytic
framework was laid out in early studies
\citep{fall1980intro,dalcanton1997intro,mo1998DI}, which have provided the
basis for most of the published semi-analytic work on the subject.  First
hydrodynamical simulations suffered from excessive loss of angular momentum,
resulting in too compact galaxies compared to the observational measurements
\citep{steinmetz1999sim,navarro2000sim}.  The origin of this problem was
identified in the excessive cooling (and star formation) during the early
phases of galaxy formation, which was due to a combination of limited numerical
resolution and stellar feedback implementation
\citep{weil1998sim,eke2000sim,abadi2003sim,governato2004sim}.  In more recent
years, many groups have succeeded in reproducing realistic thin disks supported
by rotation \citep{governato2010sim,guedes2011sim,danovich2015cold_accr,
  murante2015}.  Recent studies also focused on the origin of the specific
angular momentum versus stellar mass relation in simulations
\citep{ubler2014j_fs,marinacci2014,teklu2015magneticum_j,genel2015j_illustris,zavala2016eagle_j,defelippis2017illustris,sokolowska2017j_zoomin,grand2017j_model}.
These groups have shown that strong feedback at high redshift is crucial to
remove low angular momentum gas and produce galaxy disks with sizes and
rotational properties comparable to those measured in the local Universe
\citep{ubler2014j_fs}. Numerical work has also shown that gas dissipation
during galaxy--galaxy mergers plays an important role in determining the sizes
and angular momentum of ET galaxies
\citep{hopkins2009dissipation,Hopkins_etal_2014,
  porter2014dissipation,lagos2018EAGLE_jmerger}. This has been confirmed by a
number of studies relying on semi-analytic models of galaxy formation
\citep{shankar2014, tonini2016}. Although most modern models also include a
specific treatment for the exchange of angular momentum among the different
baryonic components
\citep[e.g.][]{lagos2009sam,guo10,benson2012sim,tonini2016,xie2017sam}, less
attention has been devoted to the  specific angular momentum of galaxies
in the semi-analytic framework (dedicated studies have been published by
\citealt{lagos2015sam_j} and \citealt{stevens2016darksage}).

In this paper, we will use a state-of-the-art semi-analytic model including
prescriptions for angular momentum evolution, to perform a systematic
analysis of both size and specific angular momentum distributions of
galaxies of different morphological type.  
The layout of the paper is as follows: in Section~\ref{sec:model}, we
give details about the semi-analytic model used in this work, and about the
dark matter simulations employed. In
Section~\ref{sec:r_ms}, we describe the size--mass relation predicted
by our model and different variants considered, and we compare it to
observational measurements both for LT and ET galaxies.  We discuss the stellar
specific angular momenta of our model galaxies in
Section~\ref{sec:ang_mom}, while in Section~\ref{sec:history} we
explain the dependence of the predicted relations on morphology, gas
content of model galaxies, and specific implementation for gas
cooling and stellar feedback. Finally, in
Section~\ref{sec:conclusions}, we discuss our results and give our
conclusions.

\section{The model}
\label{sec:model}
In this work, we take advantage of the GAlaxy Evolution and Assembly
(GAEA) semi-analytic model, described in \citet{hirschmann2015}, as updated in
\citet{xie2017sam}. This model descends from that originally published
in \citet{delucia07}, but many prescriptions have been updated
significantly over the past years.  In particular, GAEA includes a
sophisticated treatment for the non instantaneous recycling of gas,
metals, and energy \citep{delucia2014}, and a new stellar feedback
scheme partly based on results of hydrodynamical simulations
\citep{hirschmann2015}.  In this work, we use the GAEA model updated,
as described in \citet{xie2017sam}, to include a specific treatment
for angular momentum exchanges between galactic components, and
prescriptions to partition the cold gas into its molecular (star
forming) and atomic components. Specifically, we will use the model
adopting the \citet{blitz2006} prescription to estimate the molecular
gas fraction, and will refer to this implementation as X17. In this
work, we also consider several modifications of the standard X17
implementation, as detailed below. A summary of all alternative runs
considered is provided in Table~\ref{tab:runs}.

\begin{table*}
   \centering
    \caption{Summary of the model runs considered in this work. The
      second column indicates if/when gas dissipation during mergers
      is accounted for; the third column lists the proportionality
      parameter relating the specific angular momentum of the gas
      accreting in rapid mode to that of the hot gas halo; the fourth
      column indicates the adopted feedback scheme.}
    \label{tab:runs}
    \begin{tabular}{ | c | c c c |  } 		\hline
	Name	&	Dissipation & Gas accretion	&	Feedback Scheme \\ \hline
    X17     & No    &   $\alpha_{CA}=1$ &   FIRE \\
    X17allM &   During all mergers  &  $\alpha_{CA}=1$  &   FIRE \\
    X17MM   & During major mergers  &  $\alpha_{CA}=1$  &   FIRE    \\
    X17CA3  & During major mergers  &  $\alpha_{CA}=3$ &   FIRE    \\
    X17G11  & During major mergers  &  $\alpha_{CA}=1$  &   \citet{guo10} \\
    \hline
	\end{tabular}
\end{table*}

Our fiducial model is able to reproduce the observed evolution of the
galaxy stellar mass function up to $z\sim7$ and of the cosmic star
formation rate density up to $z\sim10$ \citep{fontanot2017gaea}.  In
addition, the model reproduces the measured correlation between
stellar mass/luminosity and metal content of galaxies in the local
Universe, down to the scale of Milky Way satellites
\citep{delucia2014,hirschmann2015}, and the evolution of the galaxy
mass--gas metallicity relation up to redshift $z\sim2$
\citep{hirschmann2015,xie2017sam}.  The model is, however, not without
problems: in particular, we have shown that massive galaxies tend to
form stars at higher rates than observed \citep{hirschmann2015}, and
that the model tends to under-predict the measured level of star
formation activity at high redshift \citep{xie2017sam}.

In the following, we provide a brief description of the
\citet{xie2017sam} model, focusing only on those prescriptions that
are relevant for this work.  For more details, we refer to the
original papers by \citet{delucia2014}, \citet{hirschmann2015}, and
\citet{xie2017sam}.

\subsection{The cosmological simulation and the merger tree}
\label{sec:MR_merger_tree}
The merger trees used in this work are based on the Millennium Simulation
\citep[MRI,][]{springel2005}, and on the higher resolution Millennium II
Simulation \citep[MRII,][]{boylan-kolchin2009}.  The MRI follows the evolution
of N=$2160^3$ particles of mass $8.6\times10^8 h^{-1}{\rm M_{\sun}}$, in a box
of $500\; h^{-1}$Mpc comoving on a side.  Simulation outputs are stored in 64
snapshots, logarithmically spaced in redshift.  The cosmological model adopted
is consistent with WMAP1 data \citep{spergel2003WMAP1}, with cosmological
parameters $\Omega_{b}=0.045$, $\Omega_{m}=0.25$, $\Omega _{\Lambda}=0.75$,
$H_0=100h\;{\rm Mpc^{-1}\;km\;s^{-1}}$, $h=0.73$, $\sigma_{8}=0.9$, and $n=1$.
This cosmology is nowadays out-of-date, and more precise estimates of the
cosmological parameters are available from e.g. the PLANCK collaboration
\citep{planck2014}.  In particular, the most recent estimates converge towards
a lower value of $\sigma_{8}\;(=0.829)$.  As this parameter heavily influences
the clustering of cosmic structures, a different value of $\sigma_8$ is
expected to affect also the evolution of model galaxies.  Previous studies
\citep{wang_jie_2008,guo_2013_wmap} have shown, however, that model results are
qualitatively similar when run on a simulation with lower $\sigma_8$, although
the different cosmology requires a slight retuning of the physical parameters
of the model.  We do not attempt here to rescale the cosmology of the MR
  as done, for example, in \citet{guo_2013_wmap}. Based on previous results, we
  expect that such modifications would not affect significantly our results.

The halo merger trees used as input for our galaxy formation model are
built in different steps.  First, halos and sub-halos are identified
for each snapshot. Halos are identified using a classical
Friends-of-Friends algorithm, with a linking length equal to 0.2 times
the mean inter-particle separation. The SUBFIND algorithm
\citep{springel2001accr_mode} is then used to identify bound
substructures in each FoF halo. As in previous work, we consider as
genuine substructures only those with at least 20 particles, which
sets the halo mass resolution to $M_{h}=1.7\times 10^{10}
  h^{-1}{\rm M_{\sun}}$ for the MRI.  For each subhalo at any given
snapshot, a unique descendant is identified at the subsequent
snapshot, by tracing a subset of the most bound particles
\citep{springel2005}.  In this way, each subhalo is automatically
linked to all its progenitors (at the previous snapshot), i.e. its
merger history is defined.  For each halo, a main branch is defined as
the one that follows, at each node of the tree, the progenitor with
the largest integrated mass \citep{delucia07}.

When necessary, to verify the robustness of our results at masses near
the resolution limit of the simulation, we use the MRII.  This
corresponds to a simulation box with size of $100\; h^{-1}$Mpc on
a side (one fifth of the Millennium), with a particle mass that is 125
times smaller than that used in the MRI. This lowers the halo mass
resolution to $M_{\rm h}=1.4\times10^8h^{-1}{\rm M_{\sun}}$. The resolution
limits of the MRI and MRII simulations translate in stellar mass
limits for the X17 model of about $\sim 10^9\;{\rm M_{\sun}}$ and
$\sim 10^8\;{\rm M_{\sun}}$ for the MRI and MRII, respectively
\citep[see Fig.~6 of ][]{xie2017sam}.

\subsection{The fiducial semi-analytic model}
Our semi-analytic model attaches baryonic components to each simulated
halo, considering their merger histories and observationally and/or
theoretically motivated prescriptions to model the evolution of the
baryons. In this section, we focus on the processes driving the
evolution of galaxy sizes and angular momenta, i.e. the main subject
of this work.

\subsubsection{Cooling}
When a halo collapses, it is assigned a hot gas component, and the total 
baryonic mass in the halo is assumed to be
$M_{baryons}=f_bM_{200}$ ($f_b$ is the universal baryon fraction, and
$M_{200}$ is defined as the mass corresponding to an over-density of
200 times the critical density of the Universe).  In our model, the
hot gas is assumed to follow an isothermal distribution and can cool
only onto central galaxies. The process is modeled as described in
detail in \citet{delucia2010cooling}, following the original
prescriptions outlined in \citet{white1991SAM}: a cooling radius is
defined as the radius at which the local cooling time is equal to the
halo dynamical time. Two different cooling regimes are considered,
depending on how the cooling radius compares to the virial radius
$R_{200}$ (the radius corresponding to $M_{200}$). At high redshift
and for small haloes, the formal cooling radius is much larger than
the virial radius.  In this case, the infalling gas is not expected to
reach hydrostatic equilibrium. Gas accretion is anisotropic
(filamentary) and limited by the infall rate.  In this `rapid cooling
regime' (or `cold accretion mode'), we assume that all the hot gas
available cools in one code time step.  When instead the cooling radius is smaller than the
halo virial radius, the hot gas is assumed to reach hydrostatic
equilibrium and to cool quasi-statically. In this `slow cooling
regime' (or `hot accretion mode'), the cooling rate is modeled by a
simple inflow equation.

In both regimes, we assume that the hot gas transfers angular momentum
to the cold gas disk, proportionally to the cooled mass,
$M_{\rm cooling}$.  As in previously published models, the
hot halo is assumed to have the same specific angular momentum as the
dark matter halo $\vec{j}_{\rm DM}$, so that the specific angular momentum
of the cold gas after cooling can be written as:
\begin{equation}
\label{eq:j_cooling}
\vec{j}_{\rm cold}^f = \frac{\vec{j}_{\rm cold}^0 M_{\rm cold}^0 + \alpha_{A}\,\vec{j}_{\rm DM} M_{\rm cooling}}{M_{\rm cold}^0+M_{\rm cooling}} . 
\end{equation}
In the above equation, $\vec{j}_{\rm cold}^f$ and $\vec{j}_{\rm cold}^0$ are
the specific angular momenta of the cold gas after and before gas
cooling, $M_{\rm cold}^0$ is the mass of the cold gas disk before cooling,
and $\alpha_{A}$ is assumed to be 1 in our fiducial model.  Several
recent studies based on hydrodynamical simulations
\citep{stewart2011cold_accr,pichon2011cold_accr,danovich2015cold_accr}
have shown that, contrary to standard assumptions, gas accreted
through cold mode carries a specific angular momentum that is from 2 to 4
times larger than that of the parent DM halo. To quantify the
influence of this effect on our model results, we have carried out an
alternative run assuming $\alpha_{A}=3$ for gas accreted in the rapid
cooling regime. We will refer to this run, which adopts the same model
parameters as the reference X17 model, as X17CA3 in the following.

\subsubsection{Star formation and stellar feedback}
\citet{xie2017sam} introduced modified prescriptions for the star
formation process, aimed at including an explicit treatment of the
atomic-to-molecular gas transition. In our reference X17 model, the
total cold gas reservoir associated with each galaxy is partitioned
into its molecular and atomic components employing the
\citet{blitz2006} empirical relation. The star formation rate is then
assumed to depend on the surface density of molecular gas in the disk
\citep{wong2002sf,kennicutt2007sf,leroy2008}.

We assume that the newly formed stars, $M_{*,new}$, carry the angular
momentum of the cold gas they originated from. Therefore, after a star
formation episode, the specific angular momentum of the stellar disk,
$\vec{j}_{*,disk}^f$, can be written as:
\begin{equation}
\vec{j}_{\rm *,disk}^f = \frac{\vec{j}_{\rm *,disk}^0 M_{\rm *,disk}^0 + \vec{j}_{\rm cold} M_{\rm *,new}}{M_{\rm *,disk}^0+M_{\rm *,new}},
\end{equation}
with $\vec{j}_{\rm *,disk}^0$ and $M_{\rm *,disk}^0$ representing the specific
angular momentum and stellar mass of the disk before star formation,
and $\vec{j}_{\rm cold}$ the specific angular momentum of the cold gas
disk at the time of star formation. Gas recycled from
stars is later returned to the cold gas, carrying the specific angular
momentum of the stellar disk or, in the case of gas originating from
bulge stars, a zero specific angular momentum\footnote{We note that in
  \citet{xie2017sam}, we assumed that gas recycled from bulge stars
  also carries the same specific angular momentum of the stellar
  disk.}.

Stellar feedback injects energy into the interstellar medium,
reheating part of the cold gas.  In our reference X17 run, the
reheating is modeled using parametrizations based on the FIRE
hydrodynamical simulations \citep{Hopkins_etal_2014,muratov2015FIRE}.
Part of the reheated gas is eventually ejected from the parent halo
through galactic winds. The amount of gas ejected is estimated using
energy conservation arguments.  We assume that reheating and/or
ejection does not affect the specific angular momentum of the cold gas
and that of the hot gas (the latter is always equal to that of the
parent dark matter halo).  The ejected gas is stored in a reservoir,
from where it can be re-incorporated onto the hot gas associated with
the parent halo, on a time-scale that depends on the virial mass of
the halo \citep[for a detailed description of the prescriptions
  adopted for our fiducial stellar feedback scheme,
  see][]{hirschmann2015}.  

Recent numerical work \citep{ubler2014j_fs,genel2015j_illustris} has
highlighted that gas ejected through galactic winds can be
accelerated, so that a larger angular momentum is transferred to the
disk when the gas is re-accreted. In our model, the ejected gas is not
accelerated, but we assume it acquires the same specific angular
momentum of the parent halo (which typically increases with increasing
cosmic time) before being re-incorporated. To quantify how much our
results depend on the feedback scheme adopted, we also consider a
different implementation based on the feedback scheme used in
\citet{guo10}. For this particular prescription, we have used the same
parameters adopted in \citet{hirschmann2015}, and have not attempted
to re-tune them. As shown in our previous work, this feedback scheme
does not reproduce, in our GAEA framework, the measured evolution of
the galaxy stellar mass function, and implies lower ejection rates of
gas at high redshift and shorter re-accretion times with respect to our
reference X17 model. This different re-accretion history is expected
to affect significantly the star formation history of model galaxies,
and therefore also their sizes and angular momenta.  In the following,
we refer to the run adopting the \citet{guo10} stellar feedback
parametrization as X17G11.

\subsubsection{Bulge formation}
\label{sec:bulge_formation}
Mergers and disk instabilities are the two possible channels that in
our model lead to the formation of a bulge. We assume this is a
spheroidal component with zero angular momentum, and supported by velocity
dispersion.

We distinguish between two types of mergers, based on the baryonic
(stars+cold gas) mass ratio between the secondary (less massive) and
primary (more massive) merging galaxies.  If the mass ratio is larger
than 0.3, we assume we have a `major merger' event during which both
stellar components of merging galaxies merge into a single remnant
bulge. In the case of a minor merger (mass ratio less than 0.3), we assume
that the stellar disk of the primary is unperturbed, and that the
stars of the secondary are added to the primary bulge. 
During all
mergers, the cold gas of the secondary is added to the cold gas disk
of the primary. We assume that the cold gas is first stripped from the
satellite, and acquires the same specific angular momentum of the
primary dark matter halo, obtaining an equation similar to that used
for cooling (Eq.~\ref{eq:j_cooling}), but with the secondary cold gas
mass instead of $M_{cooling}$.

We assume that all mergers trigger a star burst in the cold gas disk
of the remnant, which is modeled following the `collisional starburst'
prescription introduced by \citet{somerville2001}, with coefficients
revised using results from \citet{cox2008SBcoeff}.  The amount of new
stars formed, $M_{*,SB}$, is a fraction of the cold gas of the
progenitors, proportional to the merger mass ratio.

Disk instability is modeled as described in detail in \citet[][see
  also \citealt{delucia2011BulgeMerger}]{croton2006}.  The instability
criterion is based on results by \citet{efstathiou1982DI}.  When a
disk becomes unstable, a fraction of stars $\delta M$, necessary to
restore the stability, is moved from the central regions of the disk
into the bulge.  During a disk instability episode, we assume that the
angular momentum of the stellar disk is preserved, and thus the
specific angular momentum of the disk can be written as:
\begin{equation}
j_*^f = \frac{j_*^0 M_{*,disk}^0}{M_{*,disk}^0-\delta M}           
\end{equation}
with $M_{*,disk}^0$ representing the initial mass of the disk
before the disk instability event. 

As discussed in previous work
\citep[see][]{athanassoula2008DI,benson2010DI,delucia2011BulgeMerger,
  fontanot2011bulge}, our modeling of disk instability is rather simplified:
       it does not account, for example, for the possibility that bar
        formation causes an inflow of gas towards the centre fuelling a
        starburst, or for violent early disk instability
        \citep{elmegreen2008DI,ceverino2015DI}.  In addition, the very same
      instability criterion adopted has been criticized by
      e.g. \citet{athanassoula2008DI}.  Improving the modeling adopted for this
      physical process is highly needed, but goes beyond the aims of this work.
 
\subsubsection{The disk radius and bulge size}
\label{sec:sam_sizes}
In our reference model, the radii of the cold gas and stellar disks
are estimated from their specific angular momentum and rotational
velocity.  Specifically, the disk scale
radius is expressed as:
\begin{equation}
 R_{x} = \frac{j_x}{2V_{max}},
\end{equation}
where $R_x$ and $j_x$ are the radius and the specific angular
momentum of the $x$-component (either cold gas or stellar
disk).  $V_{max}$ is the maximum rotational velocity of the parent
halo.

The bulge is assumed to be a dispersion dominated spheroid, and its
size is estimated from energy conservation arguments.  During mergers
of spheroids, the energies involved are those due to their
gravitational potential and interaction.  Assuming no energy
dissipation, we can estimate the energy before the merger as:
\begin{equation}
\label{eq:energy_1}
\begin{split}
E_i = CG \left[ \frac{(M_*^p+M_{*,SB}^p)^2}{R_p} + \frac{(M_*^s+M_{*,SB}^s)^2}{R_s} \right] \\
+ \alpha G \frac{(M_*^p+M_{*,SB}^p)(M_*^s+M_{*,SB}^s)}{R_p+R_s},
\end{split}
\end{equation}
and after the merger as:
\begin{equation}
\label{eq:energy_2}
 E_f = C G \frac{{M_*^f}^2}{R^f} .
\end{equation}

In the first equation, $G$ is the gravitational constant,
$M_*^p+M_{*,SB}^p$ and $M_*^s+M_{*,SB}^s$ are the total stellar masses
of the primary ($p$) and the secondary ($s$) galaxy, respectively.
$M_{*,SB}$ represents the stars formed during the starburst, and $R_p$
and $R_s$ are approximations of the half mass radii of the primary and
secondary galaxies (including cold gas). The latter are obtained from
a mass-weighted average of the sizes of the bulge and disk components.
$C=0.5$ is a `form factor', and $\alpha=0.5$ is a parameter accounting
for the orbital energy between the spheroids \citep{cole2000}.  In the
last equation, $M_*^f$ is the stellar mass of the remnant spheroid,
and $R_f$ is its size.

Previous studies have highlighted how this simple treatment leads to
unrealistic sizes of galaxies, especially at the low mass end
\citep{hopkins2009dissipation,covington2011dissipation,porter2014dissipation}.
This problem arises from the fact that the model outlined above
ignores gas dissipation in bulge formation through mergers.  Using
high-resolution hydro-simulations, \citet{hopkins2009dissipation}
proposed a simple formula to account for gas dissipation, without
modifying the energy conservation equation. The final radius can be
simply `corrected' as follows:
\begin{equation}
\label{eq:rdiss}
 r_{f} = \frac{r_{no\, diss}}{1+f_{gas}/f_0}
\end{equation}
where $r_{no\, diss}$ represents the bulge size when dissipation is
not considered, $f_{gas}$ is the ratio between the gas and the stellar
mass involved in the merger (including the stars formed in the
associated star-burst), and $f_0$ is a parameter varying between
$0.25$ and $0.30$.
This formula was calculated using a set of controlled simulations of
binary mergers with mass ratio larger than 1:6, and the strongest
effect was found in the case of disk--disk major mergers \citep[see for
  example Table 1 in ][]{porter2014dissipation}.  Therefore, it is not
straightforward to apply this correction to all mergers. To understand
the impact on our model results, we consider two alternative
implementations: we either assume that gas dissipation affects bulge
size during {\it all} mergers (this run is referred to as X17allM in
the following), or we assume that gas dissipation matters only during
major mergers (X17MM). In both runs, we assume $f_0=0.275$.  We have
not distinguished mergers between disks from mergers between spheroids (we
expect that mergers between spheroids involve typically a small
fraction of gas).  We also tested an alternative implementation for
gas dissipation,  which includes a dissipation term in the energy
conservation equation during mergers, as proposed by
\citet{covington2011dissipation}.  The results obtained using this
alternative implementation are qualitatively similar to those obtained
using Eq.~\ref{eq:rdiss}.

 When a disk instability episode occurs, we compute the radius $R_{\delta
    M}$ enclosing the stellar mass moved from the central part of the disk to
  the bulge (this is done assuming a disk with an exponential surface density).
  We then use this radius as the scale-radius of the newly formed spheroidal
  component.  If a bulge already exists, we merge the newly formed spheroid
  with the pre-existing bulge, assuming energy conservation, as in
  Eqs.~\ref{eq:energy_1} and \ref{eq:energy_2}.


\section{The size-mass relation}
\label{sec:r_ms}

\begin{figure*}
    \includegraphics[trim=2cm 0.8cm 2.5cm 2cm, clip, width = 0.99\linewidth]{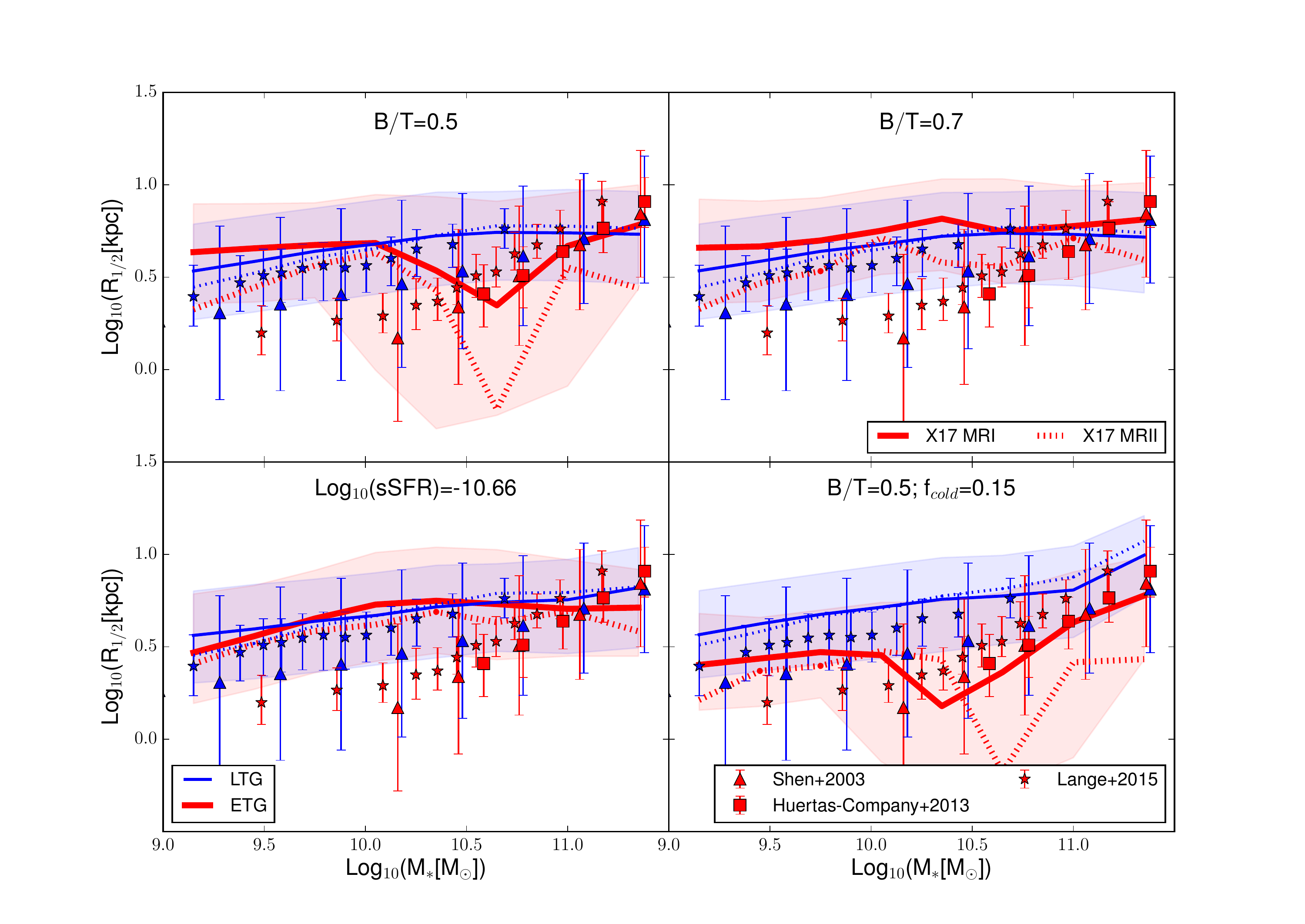}
    \caption{The $R_{1/2}$--$M_*$ relation for LT and ET galaxies (blue
      and red lines) for our reference model X17. Solid lines
      correspond to results from the MRI, while dashed lines
      correspond to the MRII.  Different panels show different
      selections for LT and ET galaxies.  Shaded areas show the region
      between the 16th and 84th percentiles of the MRI distribution;
      we find a similar scatter for the MRII.  Symbols with error bars
      correspond to different observational measurements, as indicated
      in the legend.}
    \label{fig:reff_ms_std}
\end{figure*}

In this section, we study the size--mass relation for model galaxies
divided into LT and ET galaxies, and compare model predictions to
available observational measurements in the Local Universe.  The
results of this comparison are shown in Fig.~\ref{fig:reff_ms_std},
for different LT/ET selections (different panels).  Different colors
correspond to different galaxy types (red for ET and blue for LT
galaxies), while different line styles are used for model results
based on the MRI (solid) and on the MRII (dotted lines).  The shaded
areas indicate the region between the 16th and 84th percentiles of the
distribution obtained for the MRI, but a similar scatter is found for
the MRII.  The sizes shown in the figure correspond to the projected half-mass
radii of model galaxies, namely the radii enclosing half of the total
stellar mass. To estimate radii for our model galaxies, we assume that
all galaxies are seen face-on, an exponential profile for the stars in
the disk, and a Jaffe profile for bulge stars (see Appendix
\ref{app:profiles}).

Observational estimates are shown in Fig.~\ref{fig:reff_ms_std} as symbols with
error bars (red and blue are used for ET and LT galaxies, respectively).  All
the estimates shown correspond to half mass radii, but are based on
observations at different wavebands, different assumptions about the light
distribution, and different selections for ET and LT galaxies.
\citet[][triangles]{shen2003_sdss} used SDSS data in the $z$-band to estimate
Petrosian half-light radii.  LT and ET galaxies were classified according to
their concentration, with E/S0 galaxies being classified as those with
concentration larger than $c=2.86$. \citet[][squares]{huertas2013size}
estimated the half-light radii performing a double component S\'ersic fitting
for all galaxies from the SDSS DR7 spectroscopic sample, and then selected 
  and studied only ET galaxies taking advantage of a machine learning
technique. \citet[][stars]{lange2015} estimated the half-light radii along the
major axis of GAMA galaxies, using single S\'ersic fits. They used elliptical
fits, and showed that these give systematically larger radii at fixed stellar
mass than the  circular fits used in previous studies.  They divided LT
from ET galaxies using four different methods: a visual morphology
classification, a S\'ersic index threshold of $n=2.5$, a color--color
$(u-r)$--$(g-i)$ division, a combination of S\'ersic index and $(u-r)$
color. They found that these different criteria select different galaxy
samples, but the size--mass relations obtained are very similar. In this work
we use their estimates based on the S\'ersic index classification.

In our comparison, we assume that light distribution traces the mass
distribution, and we do not attempt to reproduce the ET/LT classification
adopted in observational studies to avoid further assumptions.  In
Fig.~\ref{fig:reff_ms_std}, we show four different LT/ET selections based on
the predominance of the bulge components (in terms of stellar mass) and star
formation activity (which generally correlates with the amount of cold gas).
In the top panels we consider a simple selection based on the bulge over total
stellar mass ratio ($B/T$): specifically, we classify as ET galaxies all those
with $B/T > 0.5$ or $B/T > 0.7$ in the top-left and top-right panel,
respectively. In the bottom left panel, we show a selection based on the
specific Star Formation Rate (sSFR). The chosen threshold ($\log_{10}({\rm
  sSFR}\;[{\rm yr^{-1}}])=-10.66$) approximately separates the two peaks of the
sSFR distribution of our model galaxies \citep[see Fig.~8
  in][]{hirschmann2015}. As we have noted in our previous work and above, the
sSFR distribution predicted by our model does not reproduce well the observed
distribution. This problem, however, does not affect the results discussed
below qualitatively. Finally, in the bottom right panel of
Fig.~\ref{fig:reff_ms_std}, we show a selection that also considers the cold
gas fraction of model galaxies, $f_{\rm cold}=M_{\rm cold}/(M_{\rm cold}+M_*)$.
Specifically, we select as LT galaxies those with $f_{\rm cold}>0.15$ and
$B/T<0.5$, and as ET galaxies those with $f_{\rm cold}<0.15$ and $B/T>0.5$.  
   With this selection we are deliberately excluding a relevant number of
  galaxies, particularly among high mass gas-poor LT galaxies.  Nevertheless,
  we believe that this selection is useful to have a point of comparison with
  the samples of spiral gas-rich galaxies that are often identified as LT
  galaxies in observations.

We expect the MRII to provide a more precise estimate of the relation
in the stellar mass range $10^9-10^{10}\;{\rm M_{\sun}}$, where the
MRI is close to its resolution limit.  We find that the MRII median
size for galaxies with stellar mass $\sim 10^9\;{\rm M_{\sun}}$ is
about 0.2 dex lower than that based on the MRI, while predictions
based on the two simulations are in good agreement at a stellar mass
$\sim 10^{10}\;{\rm M_{\sun}}$.  At larger stellar masses, since the
MRII simulation volume is smaller than that of the MRI, predictions
based on this simulation are more noisy.  In the following, we will
show results from MRII up to a galaxy stellar mass equal to $10^{10}\;
{\rm M_{\sun}}$, and results from MRI above this limit.

The predicted size--mass relation for LT galaxies is in nice agreement with
measurements by \citet{lange2015}, independently of the selection adopted.
As mentioned above, this study adopts elliptical fits rather than
  circularized ones, and we consider their approach more physical than that adopted in previous studies.
  In contrast, for ET galaxies, the observed size--mass
relation is not well reproduced by any of the selections considered.  The two
partitions that include a cut at $B/T=0.5$ are in good agreement with data by
\citet{shen2003_sdss} and \citet{huertas2013size} for galaxy masses larger than
$M_*>10^{10.5}\;{\rm M_{\sun}}$.  For less massive galaxies, the model
significantly over-predicts galaxy sizes.  The same holds, over the entire
stellar mass range shown, for the other two selections considered.  In
particular, when selecting galaxies on the basis of the sSFR or using a cut at
$B/T=0.7$, model predictions for ET galaxies are very close to those obtained
for LT galaxies.

This problem has been noted earlier
\citep{hopkins2009dissipation,covington2011dissipation,porter2014dissipation},
and is due to the fact that our fiducial model does not include a
treatment for dissipation of energy during gas rich mergers.  

\begin{figure*}
    \includegraphics[trim=2cm 0.8cm 2.5cm 2cm, clip, width = 0.99\linewidth]{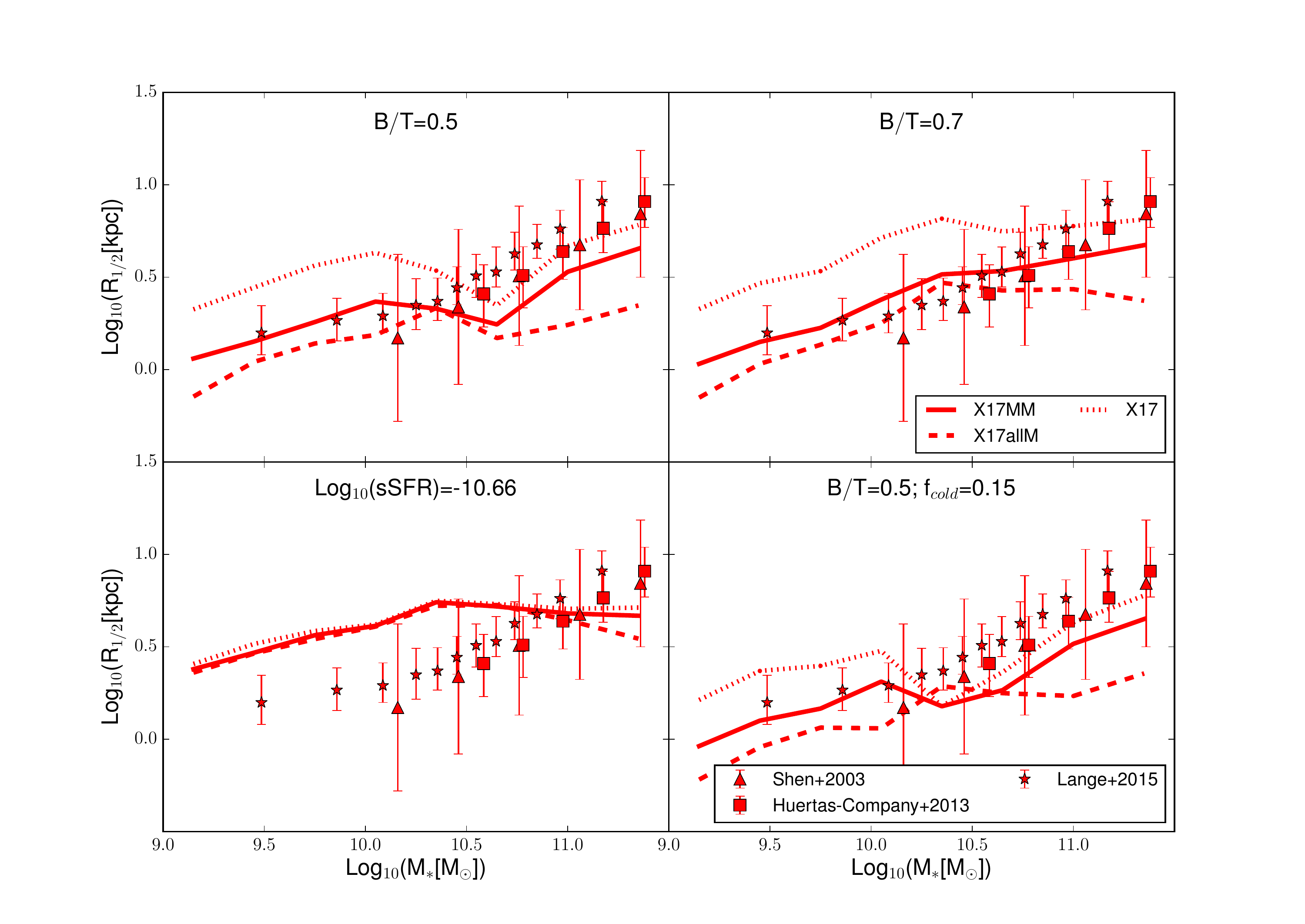}
    \caption{The $R_{1/2}$--$M_*$ relation, as in
      Fig.~\ref{fig:reff_ms_std}, but for runs including a treatment
      for gas dissipation in all mergers or in major mergers only
      (dashed and solid lines, respectively).  Predictions from our
      X17 run are shown as dotted lines, as a reference. As LT
      galaxies are not affected by the inclusion of gas dissipation,
      we only show here model predictions and data for ET galaxies.}
    \label{fig:reff_ms_diss}
\end{figure*}

In Fig.~\ref{fig:reff_ms_diss}, we show the size--mass relation for the X17allM
(dashed lines) and for the X17MM (solid lines) runs, which account for
dissipation, as in \citet{hopkins2009dissipation}.  Predictions from our
standard model without dissipation are also shown as a reference (dotted
lines).  LT galaxies are not affected by the inclusion of dissipation, and we
do not show their predicted size--mass relation (and the corresponding data)
for clarity.  For the $B/T=0.5$ and $B/T=0.7$ selections, the inclusion of
dissipation lowers the relation, slightly in the case of X17MM and more
significantly in the case of X17allM.  At the largest masses, where the X17
model is in good agreement with data, both runs including gas dissipation, 
  in particular the X17allM run, predict smaller sizes than the observational
  estimates.  The relation for galaxies selected on the basis of their sSFR is
not affected by the introduction of a treatment for gas dissipation.  This is
due to the fact that many disky galaxies are selected in the passive ET group,
keeping the predicted size--mass relation for ET galaxies very close to that
predicted by LT galaxies. Finally, the selection including a cut based on
$f_{\rm cold}$ results in a relation similar to the division based on
$B/T=0.5$. This is expected, because ET galaxies are typically gas poor.
Interestingly, dissipation during minor mergers affects significantly the sizes
of high mass galaxies ($M_*>10^{10.2}\;{\rm M_{\sun}}$).  As we will see in
Sec.~\ref{sec:history}, this is due to the fact that these galaxies have
experienced many minor mergers during their life.

Since the X17MM run is, among all runs considered here, the one
characterized by the best agreement with observational measurements
when using selections based on $B/T$, we will adopt this run as our
reference model in the rest of the paper.

\subsection{The size of galactic components}

\begin{figure}
  \centering
    \includegraphics[trim=0cm 0cm 1cm 0.5cm, clip, width = 0.9\columnwidth]{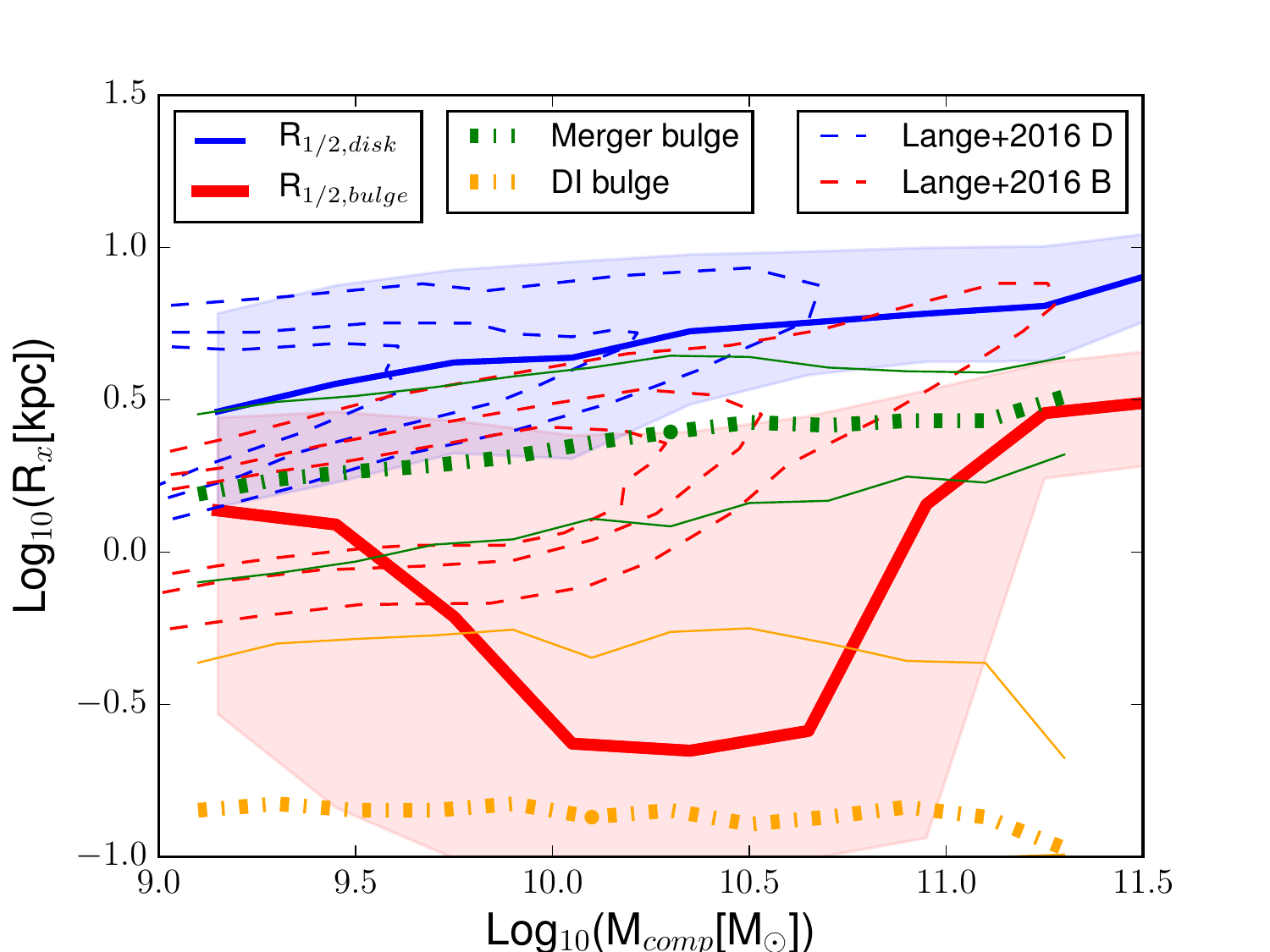}
    \caption{The size--mass relation for the bulge and disk components. 
        The x-axis represents the mass of each component.  The thick solid blue
      and red lines show the median sizes of the stellar disks and of the
      bulges of model galaxies from our X17MM run. The shaded areas correspond
      to the region between the 16th and 84th percentiles of the distributions.
      The green and orange thick  dashed-dotted lines show the median
      relations for merger and disk instability dominated bulges, respectively
      (see text for details).  Thin  solid lines of the same colors
      correspond to the 16th-84th percentiles of the distribution.  Finally,
      the dashed contours show the distributions of observational measurements
      from \citet{lange2016}, for disks (blue) and spheroids (red).}
    \label{fig:reff_components}
\end{figure}

In this section, we analyze the size--mass relation for the disk and bulge
components separately. Fig.~\ref{fig:reff_components} shows the half-mass radii
versus stellar mass of disks and bulges from the X17MM model, considering
  all the galaxies.  We also show the observed distributions by
\citet{lange2016} as dashed contours (blue for disks and red for spheroids).
The median size--mass relation predicted for disks by our X17MM run is in
fairly good agreement with observational estimates, while the bulge median
relation is offset about 0.5 dex below the observational estimates in the
stellar mass range $M_*\in[10^{9.8}-10^{11}]\;{\rm M_{\sun}}$.  The scatter is
large and there is a large overlap between data and model predictions, except
at the most massive end, where the model tends to under-predict bulge sizes
significantly.

To better understand the behavior of our model, we quantify the relative
contribution of mergers and disk instabilities to the mass of each bulge.  We
then divide model bulges according to the channel that contributed most to
their mass: if at least 50\% of the bulge mass formed from disk instabilities,
it is identified as DI bulge, otherwise it is classified as a merger bulge.
The division of bulges into two classes is motivated by observational results
 \citep[see][for a review]{kormendy2004PB}. Recent work has demonstrated
that, although characterized by different dynamical properties and radial
profiles, these two different bulge families have similar sizes, with
pseudo-bulges only slightly larger than classical ones
\citep{gadotti2009_sdss_pseudo,lange2016}.  The predicted size--mass relations
for DI and merger bulges are shown in Fig.~\ref{fig:reff_components} as green
and orange solid lines, respectively (thick lines are for the median and thin
lines for the 16th-84th percentiles of the distributions).  We find that merger
bulges are systematically larger than DI bulges, and that their median
size--mass relation is only slightly flatter than the observed distribution,
especially at high masses. Assuming that all bulges forming primarily
  through disk instability are `pseudo-bulges', this result is in stark
contrast with observational measurements.  We will come back to this issue in
Sec.~\ref{sec:discussion_di}.

\subsection{Early type central and satellite galaxies}
\label{sec:r_ms_cen_sat}
\begin{figure*}
  \includegraphics[trim=2cm 0.8cm 2.5cm 2cm, clip, width =
  \linewidth]{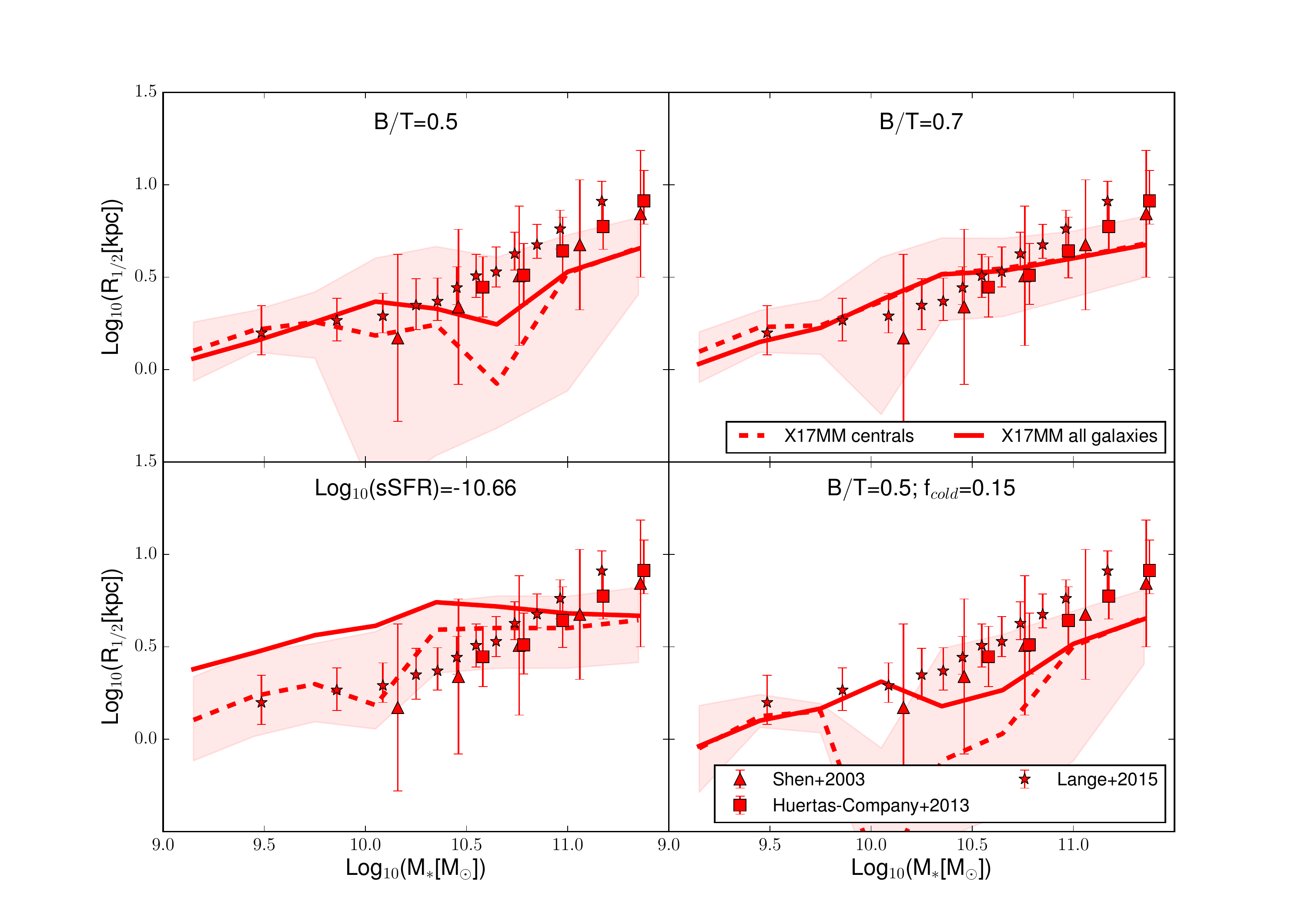}
  \caption{The $R_{1/2}$--$M_*$ relation for ET galaxies from the X17MM
    model.  Predictions for all ET galaxies are shown as solid lines,
    while dashed lines show the average sizes of central ET galaxies
    only.  The shaded areas cover the region enclosed by the 16th and
    the 84th percentiles of the central galaxies distribution. }
    \label{fig:reff_ms_diss_ty0}
  
\end{figure*}

Observational studies suggest that central and satellite ET galaxies
  follow the same size--mass relation, at least at the high mass end
  \citep[$M_*>10^{10.5}\;{\rm M_{\sun}}$, see for example][]{huertas2013size}.
In Fig.~\ref{fig:reff_ms_diss_ty0}, we show the median size--mass relation for
all ET galaxies (solid lines) and ET central galaxies only (dashed lines) from
the reference X17MM run. LT centrals follow the same relation as all LT
galaxies, and we do not show them for clarity.

The relations found for ET central galaxies are different from the
relations found for all ET galaxies, for all the selections
considered, except when using a cut at $B/T=0.7$.  In the selections
assuming a cut at $B/T=0.5$, the size--mass relations predicted for
central galaxies are characterized by a strong `dip' in the stellar
mass range $M_*\in[10^{10}-10^{10.8}]\;{\rm M_{\sun}}$.  This dip is
not visible when considering all ET galaxies.  As explained above,
model bulges can form through mergers or disk instabilities, with the
latter giving origin, in our model, to rather small bulges.  We have
verified that, when selecting ET galaxies using $B/T>0.7$, we select
bulges formed mainly through mergers (from 93 to 100 per cent 
depending on the stellar mass range considered). In contrast,
selecting ET galaxies with $0.5 < B/T < 0.7$, we find a significant
fraction of bulges formed through disk instability (from 24 to 91 per
cent, depending on the mass range).  Therefore, bulges formed through
disk instabilities contribute significantly to the size--mass relation
of samples selected using a $B/T=0.5$.

The results shown in Fig.~\ref{fig:reff_ms_diss_ty0} can, at least in part, be
explained by a difference in numbers between centrals and satellites in the
stellar mass range $M_*\in[10^{10}-10^{10.8}]\;{\rm M_{\sun}}$.  Specifically,
we find many more centrals in this mass range with $0.5<B/T<0.7$ than with
$B/T>0.7$. When considering the entire ET populations, the proportions are
inverted, and we find more galaxies with $B/T>0.7$.  Therefore, in this
mass range, central galaxies include a larger fraction of small bulges than the
entire ET population.  In Sec.~\ref{sec:history_DI}, we will show,
additionally, that the sizes of bulges formed through disk instabilities are
different in central and satellite galaxies, because disk instability occurs
under different conditions in these different galaxy types. As a result,
satellites have disk instability bulges larger than those of centrals by about
$\sim 0.7$ dex.

When the $B/T>0.5$ selection is combined with the $f_{\rm cold}<0.15$ cut,
the dip in the ET central galaxies size--mass relation is much more
pronounced than for the simple $B/T>0.5$ selection.  We will show in
Sec.~\ref{sec:history_gas} that gas poor galaxies and disk
instabilities are both more likely to occur in halos with low
specific angular momentum.  Therefore, a selection based on low gas
fraction likely correlates with a higher occurrence of disk
instabilities, and thus with small bulges.

Interestingly, the relation for ET central galaxies selected using a
sSFR cut is in quite good agreement with observational data.  The
entire population includes many disky quenched galaxies, but this is
not the case for central galaxies.  The reason is in the different
distributions of sSFR for centrals and satellites: all satellite
galaxies are around our threshold, $\log_{10}({\rm sSFR})=-10.66$, while
central galaxies are distributed to higher values, with a small tail
below the sSFR threshold.  Only central galaxies with large stellar
mass ($M_*\in[10^{10.8}-10^{11.5}]\;{\rm M_{\sun}}$) are found in the
low sSFR region.  Satellites are mostly quenched with respect to our
threshold because of the assumption of instantaneous hot gas
stripping.  Moreover, satellite morphology is generally preserved
after accretion, because mergers between satellites are rare, and the
only channel for bulge formation is disk instability.  Thus, star
formation and morphology in satellites are uncorrelated, while a
strong correlation is found between these two galaxy properties for
model central galaxies (i.e. quenched centrals tend to have an ET
morphology, and active centrals tend to be disk dominated).

\subsection{The size--mass relation for X17 modifications}
\label{sec:r_ms_CA3_G11}

\begin{figure*}
\centering
    \includegraphics[trim=2cm 1.cm 2.5cm 1cm, clip, width = \linewidth]{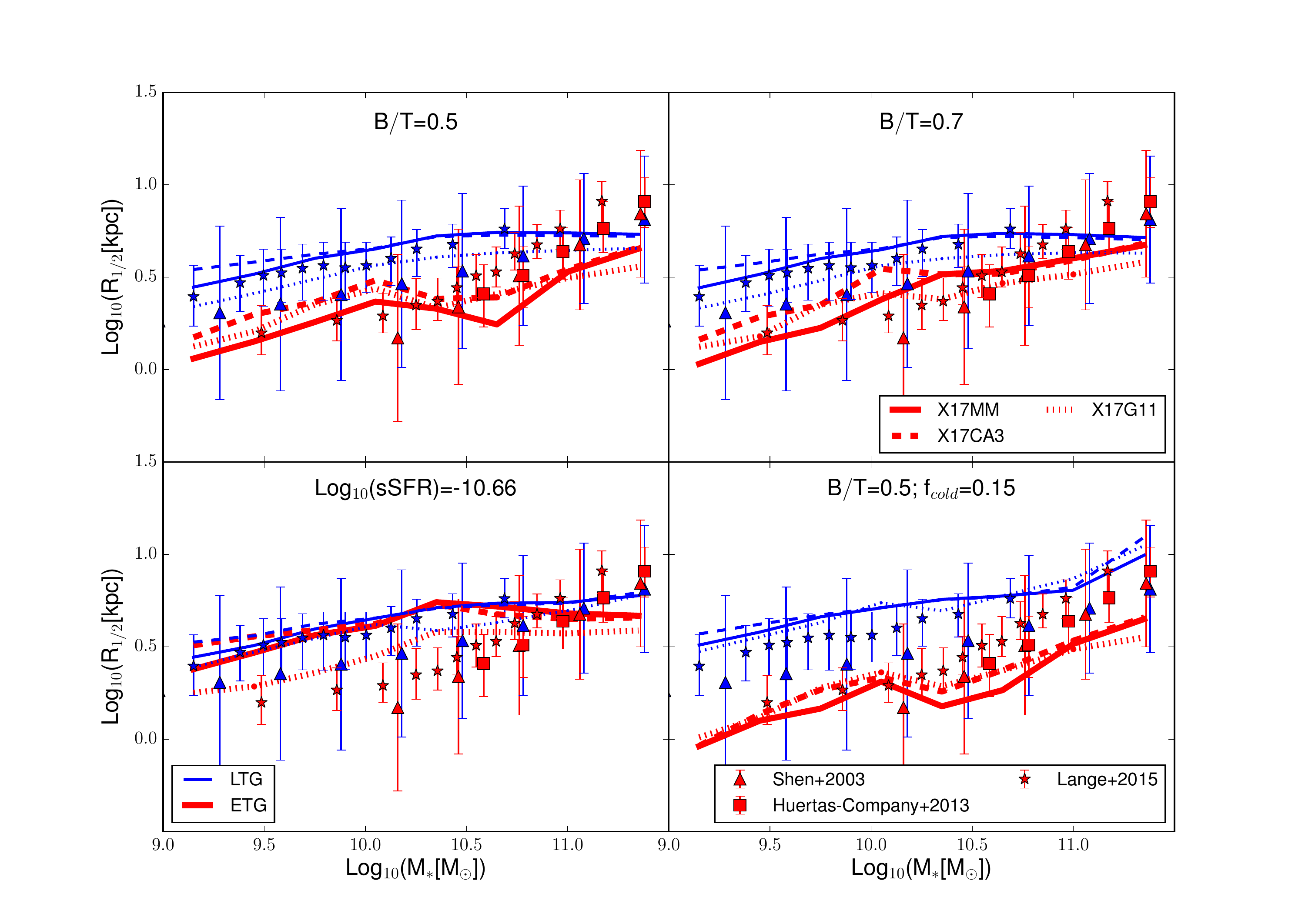}
    \caption{The $R_{1/2}$--$M_*$ relation, as in Fig.~\ref{fig:reff_ms_diss},
      for the X17CA3 (dashed lines) and X17G11 (dotted lines) runs, both
      including a treatment for dissipation during major mergers.  Predictions
      from the X17MM model are shown as solid lines, as a reference.   In
        the X17CA3 run, we assume that gas cooling through cold accretion has a
        specific angular momentum three times larger than that of the DM halo.
        In the X17G11 run, we assume an alternative stellar feedback scheme,
        based on that presented in \citet{guo10}. }
    \label{fig:reff_ms_CA3_G11}
\end{figure*}

In this section we analyze the size--mass relation for the model runs
with an increased specific angular momentum of gas during cold
accretion (X17CA3), and with a modified stellar feedback scheme
(X17G11).  We show the size--mass relations for these models in
Fig.~\ref{fig:reff_ms_CA3_G11}, with results from X17CA3 shown as
dashed lines, and X17G11 as dotted lines.  Predictions from the X17MM
model are plotted as solid lines as a reference.

Results from the X17CA3 run are very similar to those obtained from
the X17MM model for stellar masses larger than $M_*>10^{10.5}\;{\rm
  M_{\sun}}$.  For lower masses, the former model predicts sizes that
are offset high with respect to the X17MM run, by $\sim 0.1$ dex for
LT and $\sim 0.2$ dex for ET galaxies.  Therefore, the larger specific
angular momentum in cold accretion significantly affects low mass
galaxies.  We will see in Sec.~\ref{sec:history_variants} that, in
this mass range, the angular momentum of galaxies (and therefore their
size) is determined at early times (where cold accretion is
important), and is not significantly modified during subsequent
evolution.

A different stellar feedback influences the sizes of both ET and LT galaxies,
over the entire stellar mass range considered.  Specifically, we find that LT
galaxies in the X17G11 run have sizes that are systematically below those from
the X17MM model, by about 0.2 dex.  In Sec.~\ref{sec:history_variants}, we will
show explicitly the different evolution of galaxies in these two runs: the peak
of star formation in the X17G11 run occurs earlier than in X17MM, for both LT
and ET galaxies.  We then expect a lower size--mass relation also for ET
galaxies.  This is not the case: for selections based on a $B/T$ cut, ET
galaxies have, on average, sizes larger than those predicted by the X17MM run,
by about $\sim 0.2$ dex.  This is due to the fact that ET galaxies were subject
to a similar number of mergers in the two runs, but disk instabilities occurred
at earlier times in the X17G11 run.   Mergers at late times increase the
  size of bulges. In the X17MM run, this increase can be washed out by late
  disk instabilities, while these are less frequent in the X17G11 run. We
also find that fewer ET galaxies form mainly through disk instabilities in the
X17G11 run than in the X17MM run.

\section{The specific angular momentum}
\label{sec:ang_mom}
The correlation between sizes and masses of our model galaxies can be
interpreted in relation to the angular momentum treatment.  Indeed,
as discussed in Sec.~\ref{sec:model}, disk radii (both of the stellar
and of the gaseous component) are calculated from their specific
angular momenta. Below, we use results from our model to analyze the
relation between specific angular momentum and stellar mass, and its
dependence on morphology and cold gas fraction.

\subsection{Specific angular momentum estimate}
\label{sec:ang_mom_estimates}
To have a fair comparison with observational measurements, we
considered several estimates for the specific angular momentum of
model galaxies. 
In the following, we briefly describe the key quantities that we use 
in the discussion, and motivate our choices. 
The interested reader will find a more detailed description of our 
approach in Appendix~\ref{app:j_estimates}.

The angular momentum of the stellar component of a galaxy is the sum
of the angular momenta of all stars, each proportional to the product
between the distance from the center of the galaxy of the star and its
velocity.  Observations provide information on the projected stellar
luminosity, integrated in each pixel of the galaxy image. Velocity
information is inferred through spectroscopy (so these are velocities
along the line of sight), either slit spectroscopy typically along the
major axis of the galaxy, or integral field. In both cases,
measurements are performed out to a limited galactic radius, usually
corresponding to $\sim 1-2\;R_{1/2}$.

\begin{figure}
\centering
    \includegraphics[trim=0cm 17.3cm 11cm 0.0cm, clip, width = 0.85\columnwidth]{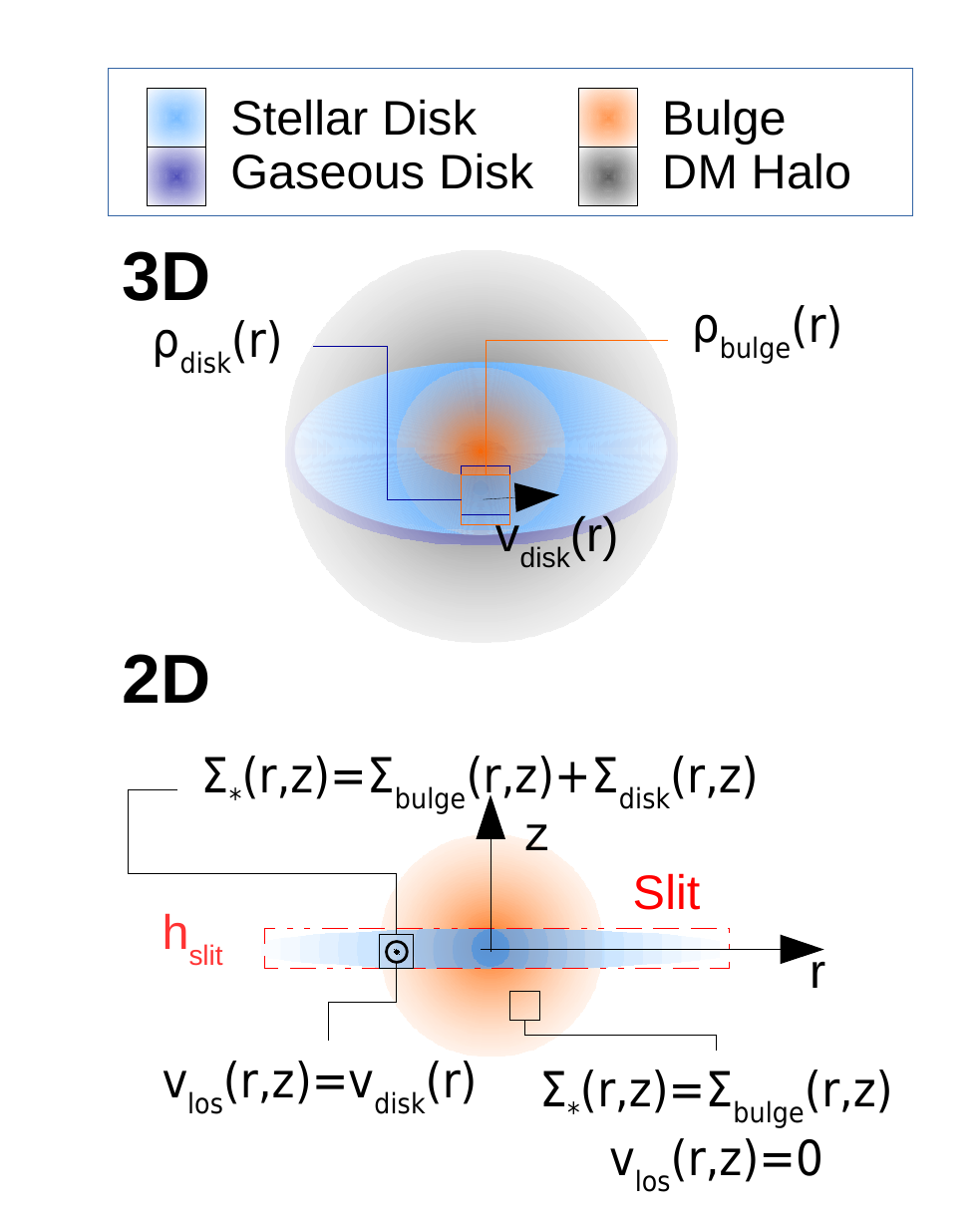}
    \caption{A schematic representation 
      of the method adopted to calculate the 3D (top) and
      2D (bottom) estimates of the angular momentum for model 
      galaxies. See text for details.}
    \label{fig:j_cartoon}
\end{figure}

To compute an estimate of the angular momentum of our model galaxies,
we assume that the stellar mass is directly proportional to the
stellar luminosity.  In addition, as for the estimate of the half-mass
radius, we assume an exponential profile for the surface density of
the rotationally supported disk and a Jaffe profile for the bulge.
Our bulge component is assumed to be dispersion supported, and
therefore has always zero angular momentum by construction.

We show a schematic representation of our typical model galaxy in
Fig.~\ref{fig:j_cartoon}.  The specific angular momentum of a galaxy can be
obtained integrating the angular momentum of its components in the 3D space. We
refer to this quantity as $j_{\rm tot}^{3D}$ in the following. Galaxies are,
however, observed in projection. In this case, the integration is performed on
the 2D plane, using the projected mass and the velocity along the line of
sight, as illustrated in the bottom panel of Fig.~\ref{fig:j_cartoon}.  We
consider two possible 2D integration methods: the first method consists in
integrating the angular momentum along a slit on the major axis of the galaxy.
In the second method, the integration is performed considering the entire
projected galaxy, in an attempt to mimic observational measurements based on
integral field spectroscopy. We refer to these two estimates as $j^{2D}_{\rm
  slit}$ and $j^{2D}_{\rm tot}$, respectively, both evaluated on model galaxies projected edge-on.  We find that our estimates
  start to converge when integrating out to $\sim 2\;R_{1/2}$ (see
  Appendix~\ref{app:j_estimates}).  In the following, when we compare our model
  predictions to observational measurements, we will adopt an integration
  radius similar to that of the observational samples considered.
 We also consider, for LT galaxies only, an estimate based on the empirical formula developed by \citet{romanowsky2012js}.
$j^{RF}$ estimates the total specific angular momentum of disk+bulge galaxies starting from their effective radius, S\'ersic index, rotational velocity at 2$R_{1/2}$ and $B/T$. 
Finally, we refer to the direct model output, corrected for $B/T$, as $j^{\rm SAM}_{\rm tot} = j^{\rm SAM}_{\rm disk} (1-B/T)$.

In the following, we will show model predictions as a shaded
area. Specifically, we will show the area between the specific angular momentum
calculated using a full 3D integration ($j^{3D}_{\rm tot}$), and that obtained
using the empirical formula by \citet[][$j^{RF}$]{romanowsky2012js} for LT galaxies. 
Typically, inclination can be easily estimated in LT galaxies, thus $j_*$ is integrated using de-projected quantities.
In the case of ET galaxies, 
inclination is rather difficult to measure, and the empirical formula does not hold. 
For these galaxies, we will show the
area between the two alternative 2D estimates: $j^{2D}_{\rm slit}$ gives an
upper limit to the estimated angular momentum, while $j^{2D}_{\rm tot}$ gives a
lower limit.
Both these measures give a lower limit for the expected relation, as our model does not include bulge rotation.

\subsection{Comparison with observations}
\label{sec:ang_mom_obs}
 The specific angular momentum of the disk, $j_{\rm disk}$, is much easier
  to estimate observationally than that of the bulge component.  In particular,
  this is straightforward when the rotational velocity profile of the disk is
  inferred from the cold gas component.  Selecting model galaxies with a cold
  gas fraction similar to that of the observational samples, we find a good
  agreement between the predicted median $j_{\rm disk}$--$M_*$ relation and
  recent observational estimates (see Appendix~\ref{app:j_disk} for more
  details). In the following, we focus on the specific angular momenta
  predicted for the entire stellar component.

\begin{figure*}
 \centering
    \includegraphics[trim=0.1cm 0cm 1cm 0cm, clip, width = 0.9\columnwidth]{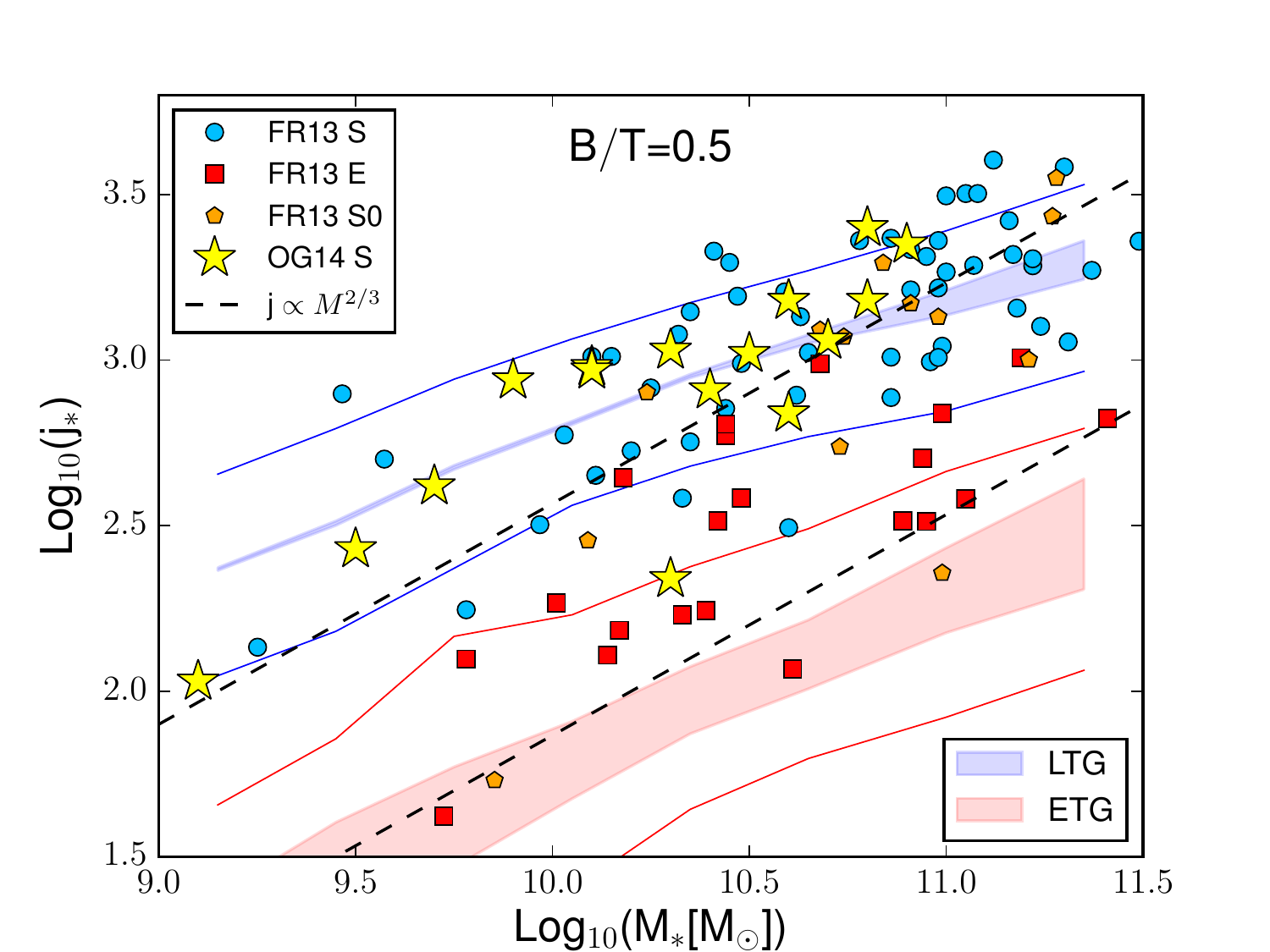}
    \includegraphics[trim=0.1cm 0cm 1cm 0cm, clip, width = 0.9\columnwidth]{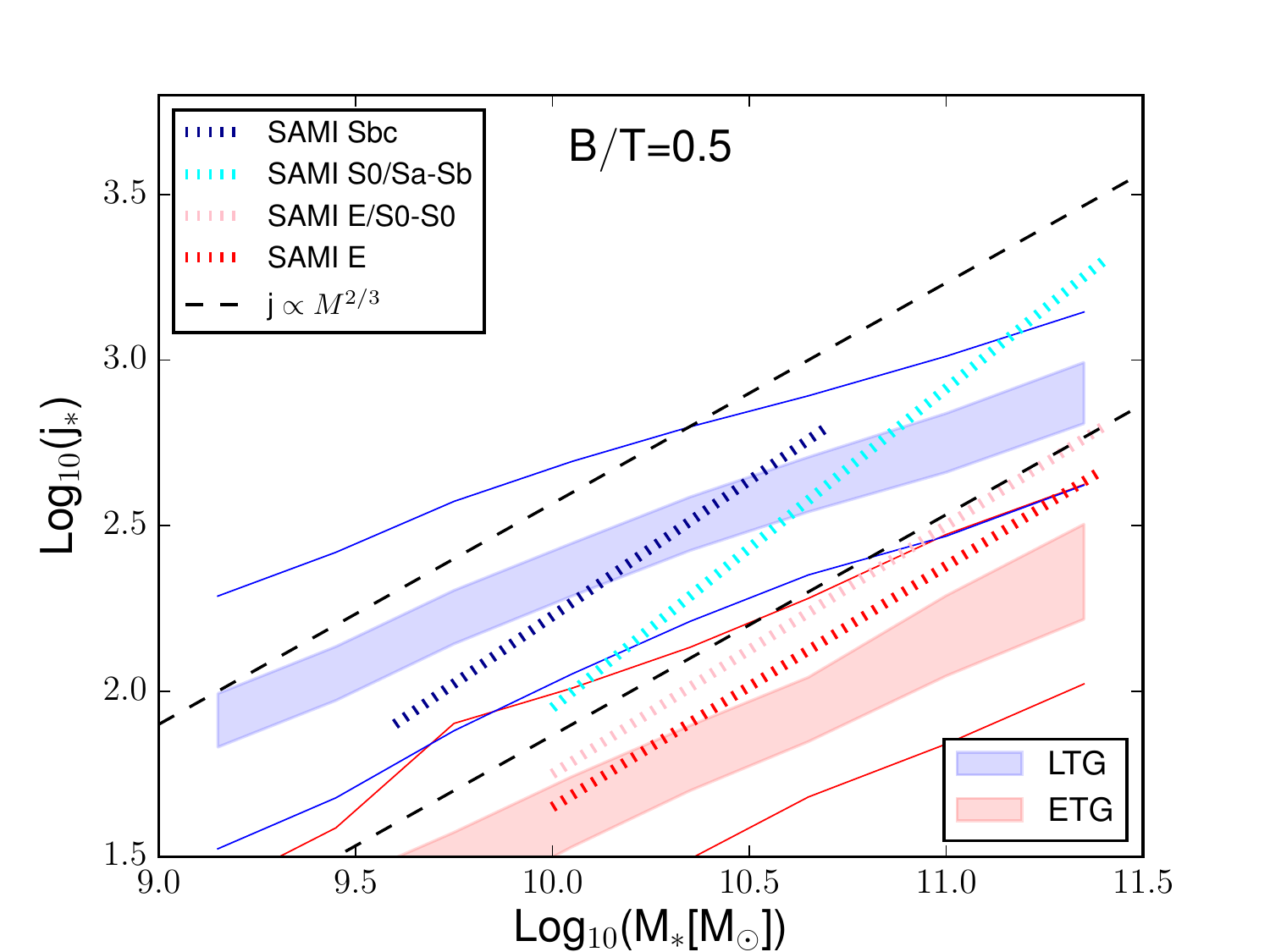}
    \caption{ The $j_*$--$M_*$ relation for LT and ET galaxies from the X17MM
      model (blue and red shaded areas, determined as described in
      Sec.~\ref{sec:ang_mom_estimates}), compared to observational data
      (symbols).  The thin solid lines correspond to the 16th-84th percentiles
      of the distributions, whereas shaded areas represent the regions covered by different estimates of $j_*$ (see text for details).  
      The radial apertures used to integrate $j_*$ in the two panels are different, and are chosen to mimic the observational data (more details in the text).
      The LT/ET selection used here assumes a
      threshold of $B/T=0.5$.  In the left panel, we show the observational
        measurements by \citet[][stars]{obreschkow2014jmbt} and by
        \citet[][other symbols]{fall2013js}, color-coded according to galaxy
        morphology: spirals (cyan), ellipticals (red) and lenticulars (orange).
        In the right panel, we show the median fits obtained for SAMI galaxies
        by \citet[][dotted lines]{cortese2016sami}.  Different colors
      correspond to different galaxy morphologies, as indicated in the legend.
      We show as a reference the expected slope for DM halos ($j\propto M^{2/3}$) as dashed black lines.  }
    \label{fig:js_ms_model}
\end{figure*}

Fig.~\ref{fig:js_ms_model} shows the predicted $j_*$--$M_*$ relation compared
with three different observational samples.  Predictions from the X17MM model
are shown as shaded areas, enclosing the region of the plane between
$j^{3D}_{\rm tot}$ and $j^{RF}$ for LT galaxies (blue), and between
$j^{2D}_{\rm slit}$ and $j^{2D}_{\rm tot}$ for ET galaxies (red).  Thin solid
lines of the same colors represent the scatter (16th-84th percentiles) of the
distributions.  We only show predictions obtained using a $B/T=0.5$ cut to
distinguish between ET and LT galaxies, as different selections give
qualitatively similar results.  
In both panels, the dashed black lines represent the theoretical expectation for the slope of the specific angular momentum versus mass relation, obtained assuming the specific angular momenta of the halo and the baryons are coupled until halo collapse \citep[$j\propto M^{2/3}$,][]{mo1998DI}. 
In the left panel, we show the
  observational measurements by \citet{romanowsky2012js}, corrected for a
  variable light-to-mass ratio as in \citet[][symbols as in the
    legend]{fall2013js}, and those by \citet[][stars]{obreschkow2014jmbt}.
  \citet{romanowsky2012js} collected a sample of galaxies of different
  morphology, with rotational velocities measured using different techniques.
  The specific angular momentum was estimated using a direct integration along
  the galaxy major axis for a sub-sample of the galaxies (out to $\sim 8\, R_{eff}$), and an empirical
  formula for the rest of the galaxies (for more details see
  Appendix~\ref{app:j_estimates}).  Estimates by \citet{obreschkow2014jmbt} are
  based on 16 gas-rich spiral galaxies from the THINGS survey
  \citep{leroy2008}, and on HI spatially resolved velocity distribution. The
  integration of the specific angular momentum was carried out assuming
  circular annuli and out to $\sim 10\,R_{1/2}$.  In this panel, we show model
  specific angular momenta for LT galaxies estimated out to $7\, R_{1/2}$, assuming that, at this distance, $j_*$ is converging to its total value.  
 For ET galaxies, we perform an integration out to $2\,R_{1/2}$, to mimic the observational limits.
  In the right panel,
  we show the observational estimates by \citet{cortese2016sami}.  This work is
  based on galaxies from the SAMI survey (dotted lines), divided according to
  their morphological type: Sbc (dark blue), S0/Sa-Sb (cyan), E/S0-S0 (pink)
  and E (red) galaxies. In this case, the specific angular momentum integration
  was performed in circular annuli out to $\sim 1\,R_{1/2}$.  For clarity, we
  show the fits to the measured relations, instead of the individual data
  points.  In this panel, the model specific angular momenta are estimated out
  to $1\, R_{1/2}$, to have a fair comparison with the SAMI sample. 
  The shaded area for LT galaxies is delimited by $j^{3D}_{\rm tot}$ and $j^{2D}_{\rm tot}$, because at $1\, R_{1/2}$ there is not enough information to calculate $j^{RF}$. 
 
We find that the  model LT and ET galaxies follow parallel relations, with a
  slope similar to that measured by \citet{fall2013js} and
  \citet{obreschkow2014jmbt}, and only slightly lower than theoretical
  expectations ($j\propto M^{1/2}$  compared to the predicted $j\propto M^{2/3}$).  The specific angular momentum of model
galaxies is lower than that estimated from these
samples, slightly in the case of LT, by $\sim 0.4$ dex for ET galaxies. Comparing model predictions to SAMI galaxies, we find that model
  predictions are slightly below observational measurements for ET galaxies,
  and exhibit a shallower relation with respect to observational data for LT
  galaxies. As explained in Sec.~\ref{sec:ang_mom_estimates}, model estimates
do not account for rotating bulges that would raise the median relation found
for ET model galaxies.  This is shown explicitly in
Appendix~\ref{app:j_estimates}, where we evaluate $j_*$ assuming that model
bulges are all rotating following empirical relations. Using this simple
assumption, the median $j_*$--$M_*$ relation is shifted up by several tenth of
dex with respect to the non rotating bulge case.  On the other hand, observed
ET galaxies by \citet{fall2013js} were selected to have a measured rotational
velocity profile.  Therefore, they represent a sample biased towards fast
rotators, and likely have a $j_*$ slightly higher than average elliptical galaxies.
Slow rotators are a small but significant fraction of the observed ET galaxy population, but in our model they represent the entire ET galaxy sample.
This is unrealistic, and we should interpret our model results for ET galaxies as a lower limit. 
SAMI ET galaxies include both fast and slow rotators, and our model predictions
are closer to the median relation measured for this sample.

\begin{figure*}
    \centering
    \includegraphics[trim=0.5cm 0.2cm 1.5cm 0.5cm, clip, width = 0.9\columnwidth]{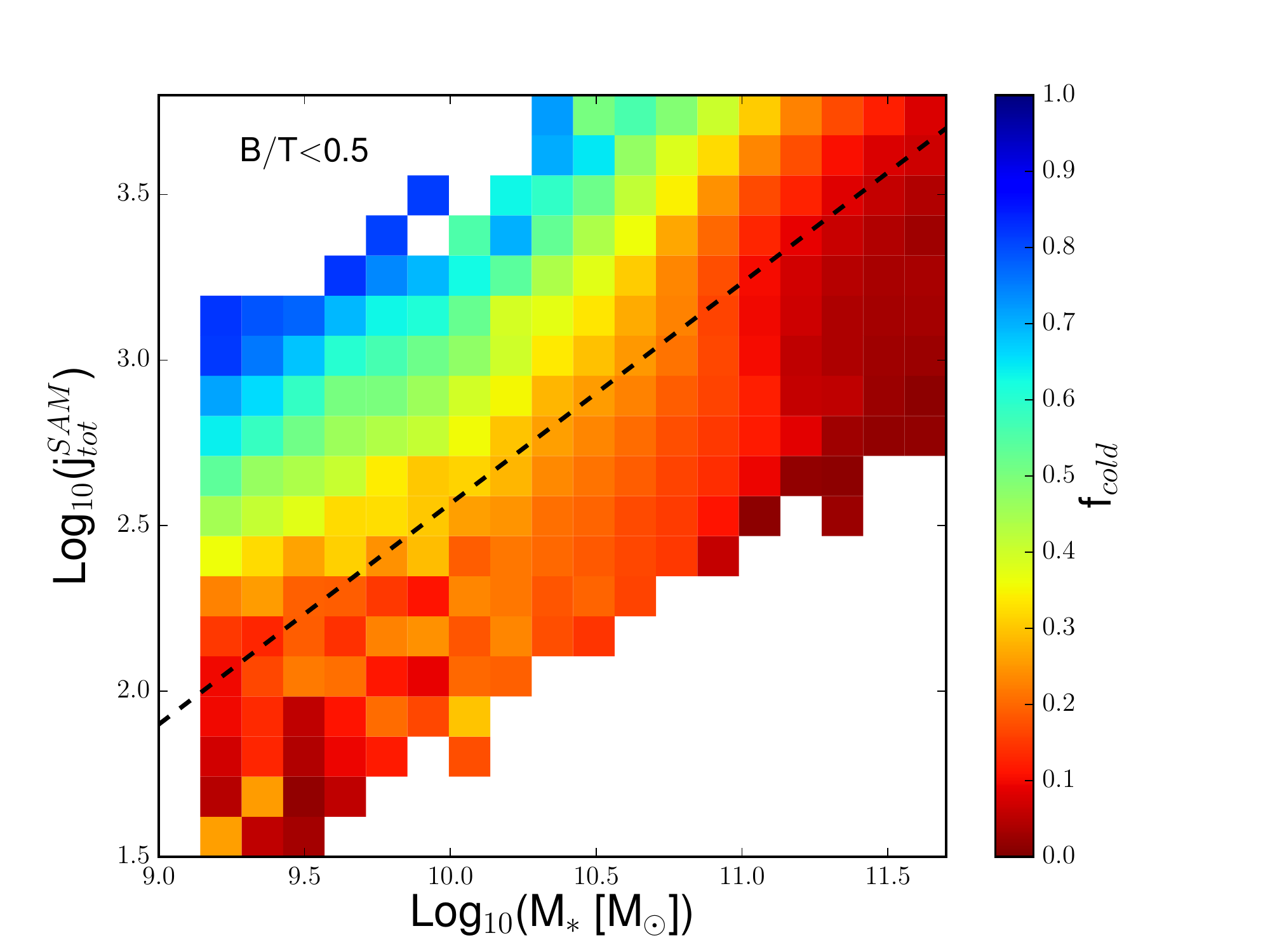}
    \includegraphics[trim=0.5cm 0.2cm 1.5cm 0.5cm, clip, width = 0.9\columnwidth]{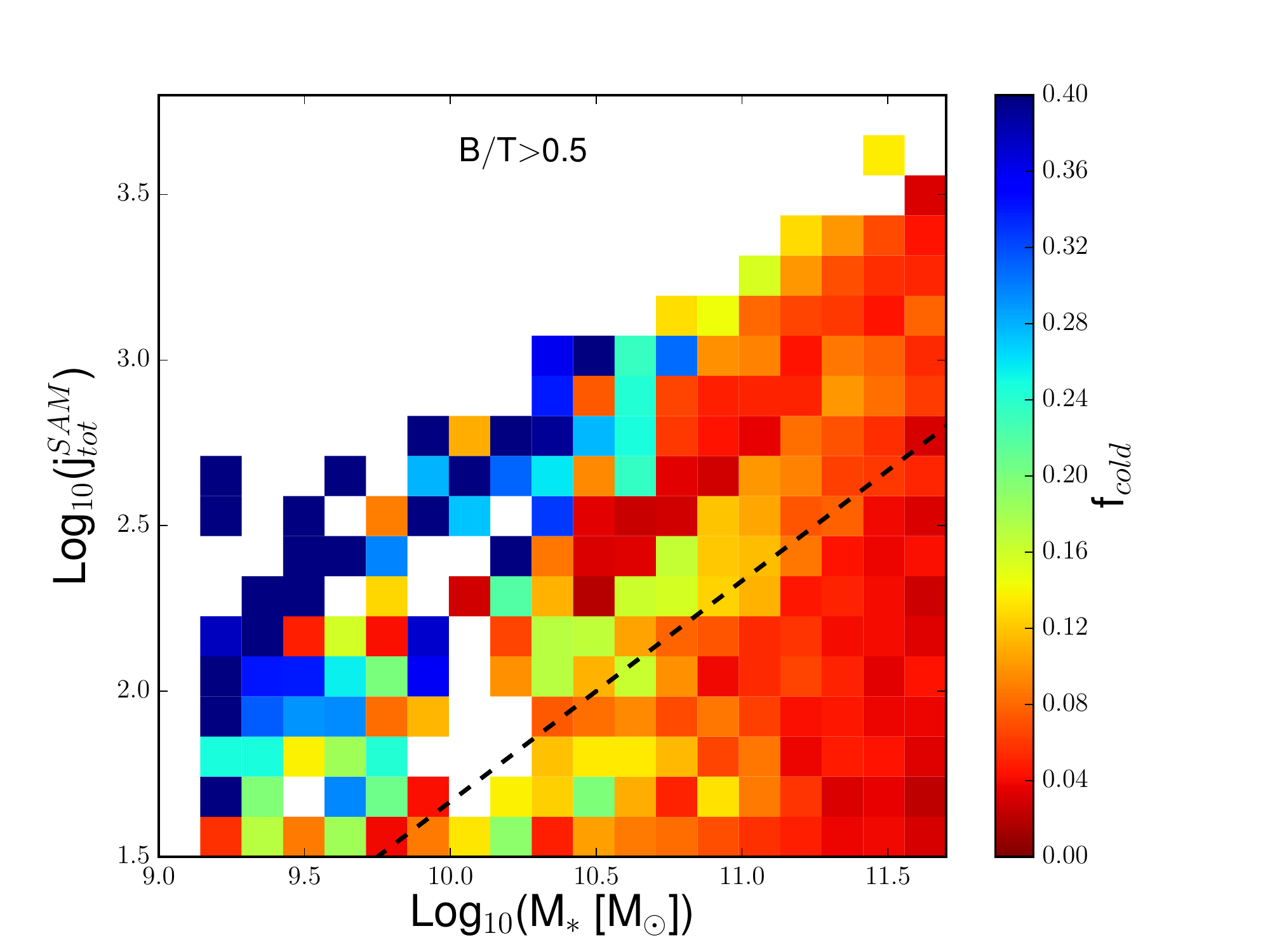}
    \caption{The $j^{\rm SAM}_{\rm tot}$--$M_*$ relation for X17MM galaxies, color
      coded by the median cold gas fraction ($f_{\rm cold}$) in each pixel.
      The left panel corresponds to LT galaxies ($B/T<0.5$), the right to ET
      galaxies ($B/T>0.5$).  A different color scale has been used in the two
      panels.   Model $j^{\rm SAM}_{\rm tot}$ values shown here correspond to the model output disk specific angular momentum, weighted for the bulge contribution.  We also show, as dashed black lines,
        the theoretical expectation for the slope of the relation ($j\propto M^{2/3}$).  }
    \label{fig:js_ms_fcold}
\end{figure*}

In the case of LT galaxies, \citet{fall2013js} and \citet{obreschkow2014jmbt}
used samples that include classical, gas-rich, spiral galaxies. Our
  selection of model LT galaxies based on $B/T$ includes both classical spirals
  and lenticulars with subdominant bulges. Therefore, compared to our model,
  observational samples appear biased towards gas-rich galaxies.  In our
model, the cold gas fraction $f_{\rm cold}=M_{\rm cold}/(M_{\rm cold}+M_*)$
strongly correlates with $j_*$, as shown in Fig.~\ref{fig:js_ms_fcold}.  
In this figure, we show the distribution of model galaxies in the $j^{\rm SAM}_{\rm tot}$--$M_*$ plane, color-coded by the median $f_{\rm cold}$ in each pixel.
We choose $j^{\rm SAM}_{\rm tot}$ as a $j_*$ estimator, in order to have a comparable quantity for LT and ET galaxies.
This quantity is not directly comparable to the shaded areas shown in Fig.~\ref{fig:js_ms_model}, but results do not change qualitatively. 
Left and right panels show the distributions of LT and ET galaxies,
respectively.  As above, we have used a simple $B/T=0.5$ cut to separate
different galaxy types, and we have used a different color scale for the two
panels, to highlight the trends as a function of the gas fractions.  Gas rich
galaxies have larger specific angular momenta than gas poor galaxies of the
same stellar mass. The relation with stellar mass is somewhat steeper for
  gas rich galaxies, and the slope in better agreement with theoretical
  expectations ($j\propto M^{2/3}$).  If we select LT galaxies with $f_{\rm
  cold}>0.3$ and $B/T<0.5$, the predicted median $j_*$--$M_*$ relation shifts
up by $\sim 0.3$ dex. 
The correlation between $j_*$ and $f_{\rm cold}$ is strong for $B/T<0.4$. 
For $B/T$ values in the range $0.4<B/T<0.7$, the correlation is not as clear, 
and there are gas-poor galaxies with quite high values of $j_*$. 
For $B/T>0.7$, the correlation is again strong. 
This effect is evident in the right panel of Fig.~\ref{fig:js_ms_fcold}, where the relation for ET galaxies exhibits a dependence on the cold gas fraction less pronounced than for LT galaxies, in particular at intermediate to high stellar masses.
This behavior can be ascribed to the presence of galaxies dominated by bulges formed mainly through disk instabilities. 
These galaxies retain, on average, less gas than galaxies with bulges formed mainly through mergers, at fixed stellar mass and $j_*$.
As we will show in detail in Sec.~\ref{sec:history_DI}, disk instabilities can form massive bulges only through a series of subsequent star formation and disk instability episodes.
Recurring star formation depletes the cold gas available in the disk, washing out the correlation between the cold gas content and the specific angular momentum.

Estimates of the angular momentum of LT galaxies based on SAMI (blue dotted
lines in the right panel of Fig.~\ref{fig:js_ms_model}) appear to follow a
steeper relation than our model predictions, and than previous observational
estimates.  As noted in \citet{cortese2016sami} their best fits are still
consistent with the expected ${2/3}$ slope. In addition, their measurements are
based on apertures smaller ($\sim 1R_{1/2}$) than those used in
\citet{fall2013js}.  For a sub-sample of their galaxies, they were able to
measure the specific angular momentum out to $\sim 2 R_{1/2}$, obtaining a
higher normalization and a better agreement with \citet{fall2013js}. When
changing the aperture radius, we find that the normalization of the
$j_*$--stellar mass relation changes, but the slope is unaffected,  as can
  be noted comparing the two panels (see also Appendix~\ref{app:j_estimates}).

\subsection{Specific angular momentum in X17CA3 and X17G11}

\begin{figure}
\centering
    \includegraphics[trim=0.1cm 0cm 1cm 0cm, clip, width = 0.9\columnwidth]{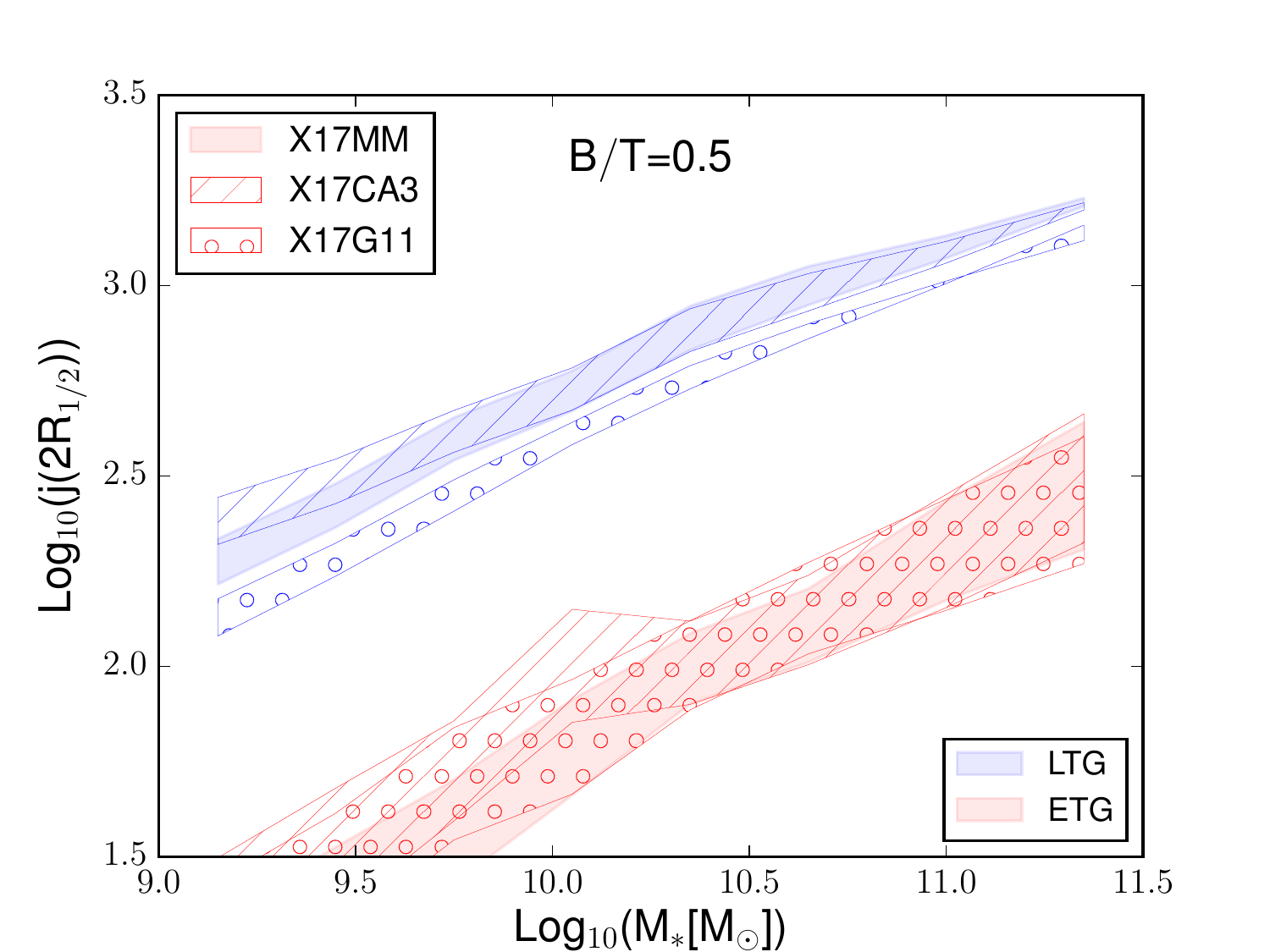}
    \caption{The $j_*$--$M_*$ relation for LT and ET galaxies (blue and
      red colors), evaluated for galaxies from the X17MM run (shaded
      areas), the X17CA3 run (hatched areas), and the X17G11 run
      (areas with circles). 
      The areas are determined as described in Sec.~\ref{sec:ang_mom_estimates}.
      All models include dissipation during
      major mergers.  LT and ET galaxies were selected using the
      $B/T=0.5$ threshold.  }
    \label{fig:js_ms_CA3_G11}
\end{figure}

We show in Fig.~\ref{fig:js_ms_CA3_G11} the $j_*$--$M_*$ relation for two
modified versions of our model: one assuming a larger angular momentum
for gas accreted during the rapid cooling regime (X17CA3, hatched
areas), and one with a stellar feedback scheme based on that used in
\citet{guo10} (X17G11, areas with circles). 
Galaxies are classified as LT and
ET using a $B/T=0.5$ cut.  We show also results from the X17MM model
as a reference (shaded areas).
The areas are determined as described in Sec.~\ref{sec:ang_mom_estimates}.
The model stellar specific angular momenta are obtained integrating $j_*$ out to $2\; R_{1/2}$, as a compromise between convergence and limited radii typically available in observational  studies.

The separation between LT and ET galaxies is evident in all the three
models considered here. The X17CA3 run returns predictions that are
almost identical to the X17MM run for massive galaxies
($M_*>10^{10.5}\;{\rm M_{\sun}}$), while at low masses the former
model predicts higher values of $j_*$ with respect to our reference
run.  Therefore, low mass galaxies are more affected by the higher
angular momentum acquired during cold accretion, while for high mass
galaxies this accretion mode is less important in determining the
final value of $j_*$.  We reached similar conclusions when analyzing
the $R_{1/2}$--$M_*$ relation.

LT galaxies in the X17G11 run have a lower median $j_*$ with respect
to X17MM, over the entire mass range considered.  
We interpret this result as due to the different stellar feedback scheme, 
which causes most of the stars to form earlier than in our reference model. 
As the angular momentum is
lower at earlier cosmic epochs, this leads to a lower normalization of
the $j_*$--$M_*$ relation.  For ET galaxies, X17G11 returns predictions
consistent with those from X17MM, for $M_*>10^{10.5}\;{\rm M_{\sun}}$.
At lower masses, the angular momentum predicted by the modified
feedback model is systematically larger than that obtained using our
fiducial X17MM run, by $\sim 0.2$ dex.  In
Sec.~\ref{sec:r_ms_CA3_G11}, we have seen that ET galaxies in the
X17G11 run have, on average, larger sizes that their counterparts of
the X17MM run. The specific angular momentum integration is influenced
by the larger size of the bulges, which translates into larger values of
$j_*$.

\section{Evolution of angular momentum and dependence on 
  other galactic properties}
\label{sec:history}
In the previous section, we have demonstrated that the scatter in the
$j_*$--$M_*$ relation correlates both with galaxy morphology and with
gas fraction.  Below, we analyze in detail the evolutionary processes
driving these correlations.

\subsection{Bulge formation channels and size--mass relation}
\label{sec:history_morphology}

In this Section, we briefly describe the different origin of LT and ET
galaxies in our model. This subject has been discussed, for previous
versions of our model, in several studies
\citep{fontanot2011bulge,delucia2011BulgeMerger,delucia2012BulgeMerger,wilman2013BulgeMerger}.
Since the basic statistics are qualitatively the same, we will focus
our analysis on those aspects that are useful to interpret the 
original results discussed in this paper. 

Our model bulge can
grow through mergers and disk instabilities.  We have evaluated the
relative importance of these two channels for galaxies in the X17MM
model, in bins of stellar mass and $B/T$. We have excluded satellite
galaxies from this analysis but most results discussed below remain
qualitatively the same when including them, except when otherwise
stated.

Table~\ref{tab:DI_mer} lists the fraction of { central} galaxies that
have experienced, from $z=1$ to the present day: (i) no relevant disk
instability event (no DI, with relevant we mean an episode
characterized by $\delta M_*/M_{*,disk}>0.1$, where $\delta M_*$ is
the fraction of stellar disk that is transferred to the bulge to restore
stability); (ii) no major merger (no MM, $M_{sat}/M_{cen}>0.3$); (iii)
no minor merger (no mM) with a mass ratio
$0.1<M_{sat}/M_{cen}<0.3$). The table shows that a significant
fraction of galaxies did not experience any minor merger since
$z=1$. This suggests that minor mergers are not the main channel for
the formation of bulges in these galaxies. Major mergers are more
likely to occur in galaxies with $B/T>0.7$, or in low mass galaxies
with $B/T>0.5$.  Disk instabilities are very frequent in the
intermediate $B/T$ bin, in particular at intermediate and high stellar
masses.


\begin{table*}
   \centering
    \caption{Fraction of {central} galaxies that did not experience a
      relevant disk instability episode (no DI), a major merger (no
      MM), or a minor merger (no mM), from $z=1$ to the present day.
      Different rows correspond to different stellar mass bins.  A
      further selection is made according to galaxy morphology:
      $B/T<0.5$ (first column), $0.5<B/T<0.7$ (second column) and
      $B/T>0.7$ (third column).
    \vspace{0.1cm}}
    \label{tab:DI_mer}
    \begin{tabular}{ | c | c c c | c c c | c c c|  } 		\hline
		&	&$B/T<0.5$	&	&	&$0.5<B/T<0.7$ 	&  	&	&$B/T>0.7$& 	\\
	$M_*\;[{\rm M_{\sun}}]\in$	& no DI	& no MM	&no mM	& no DI	& no MM	& no mM	& no DI	&no MM	&no mM	\\\hline
$[10^{9.6};\;10^{10.2}]$   & 0.97	&0.99	    &0.88	&0.76	&0.30	&0.85	&0.98	      &0.03	&0.81	\\
$[10^{10.2};\;10^{10.8}]$	& 0.94	&0.99	    &0.92	&0.12	&0.90	&0.97	&0.90	&0.09	&0.90	\\
$[10^{10.8};\;10^{11.5}]$   &0.95	&1.	        &0.82	&0.29	&0.97	&0.88	&0.96	&0.08	&0.82	\\
\hline
	\end{tabular}
\end{table*}

\begin{figure*}
\centering
    \includegraphics[trim=2.4cm 2cm 3.5cm 3cm, clip, width = 1\linewidth]{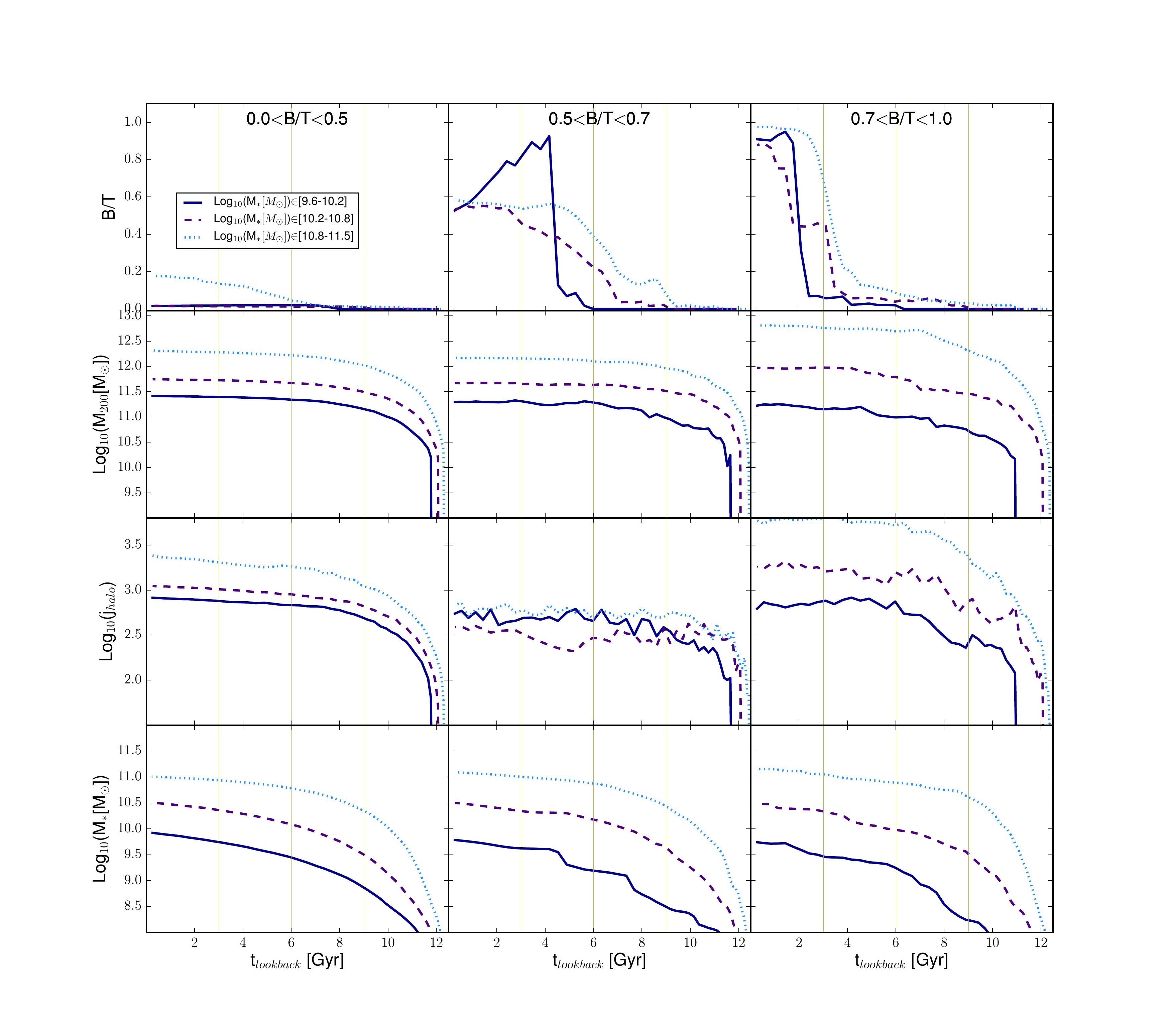}
    \caption{
     Median evolution as a function of look-back time of 
     some galactic properties, for central galaxies.
     From top to bottom: B/T, $M_{200}$, 
     $j_{\rm halo}$ and $M_*$.
     Model galaxies have been selected according to their stellar mass 
     at redshift 0: $M_* \in [10^{9.6-10.2}]$ (solid blue lines), 
     $[10^{10.2}, 10^{10.8}]$ (dashed violet lines) and 
     $[10^{10.8}, 10^{11.5}]\;{\rm M_{\sun}}$ (dotted light blue lines). 
     Different columns correspond to different values of $B/T$.
     Vertical yellow lines are for reference.
     }
    \label{fig:history_BT_med}
\end{figure*}

Fig.~\ref{fig:history_BT_med} shows the median evolution of some
selected properties as a function of lookback time.  In this figure,
central galaxies are divided according to their final stellar mass, as
in the table.  Different line-styles correspond to different stellar
mass bins as indicated in the legend, while different columns
correspond to different $B/T$ bins. The figure shows that the bulges
of low mass galaxies are formed mainly through major mergers, which
translates into an abrupt increase of the $B/T$ value following the
merger events.  Low-mass LT and ET galaxies reside in halos of
similar mass, with halos of ET galaxies only slightly less massive
than those hosting LT galaxies ($M_{200}^{LT} \sim 10^{11.4}\;{\rm
  M_{\sun}}$ and $M_{200}^{ET} \sim 10^{11.1}\;{\rm M_{\sun}}$ at
redshift 0).  On average, halos hosting ET galaxies in this stellar
mass bin formed later than those hosting LT galaxies: the former
accrete half of their final mass 9 Gyrs ago, the latter 10 Gyrs ago.
These small differences suggest that ET and LT galaxies in this mass
bin belong to the same `halo population', and that the differentiation
occurs because of the occurrence of major mergers for ET galaxies.

For the intermediate and large stellar mass galaxies
($M_*\in[10^{10.2};\,10^{11.5}]\,M_{\sun}$), we find more significant
differences between the parent halos of different types.  For
galaxies with $B/T<0.7$, the median parent halo masses at $z=0$ are
$M_{200}^{int}\sim 10^{11.7}{\rm M_{\sun}}$ and $M_{200}^{high}\sim
10^{12.3}{\rm M_{\sun}}$ for intermediate and high stellar mass bin,
respectively.  For $B/T>0.7$, the numbers become $M_{200}^{int}\sim
10^{12}{\rm M_{\sun}}$ and $M_{200}^{high}\sim 10^{12.8}{\rm
  M_{\sun}}$.  Therefore, the $B/T=0.7$ threshold separates two
different galaxy populations: one formed in relatively small halos
and the other one formed in more massive halos, that likely
experience more merger events.  Intermediate and massive galaxies with
$B/T>0.7$ form most of their bulge mass through mergers: the fraction
of galaxies that did not experience any merger in this $B/T$ range
varies between 3 and 7 per cent, depending on the stellar mass bin.
Intermediate and massive galaxies with $B/T<0.7$, in contrast, form
their bulge mainly through disk instability. In this case, the
probability of building a relevant bulge depends on the specific
history of the galaxy and of its halo.  The main difference between
galaxies with $B/T<0.5$ and those with $0.5<B/T<0.7$ is in the
specific angular momentum of their halos, $j_{\rm h}$.  Galaxies with a
more prominent bulge have a smaller $j_{\rm h}$ for most of their history.
The small $j_{\rm h}$ is transferred to the cold gas disk through cooling,
and then to the stellar disk through star formation.  Stellar disks in
galaxies with $0.5<B/T<0.7$ are thus smaller than those associated
with $B/T<0.5$ galaxies.  This affects the stability of the disk: at
fixed stellar mass, halos with smaller $j_{\rm h}$ have a higher probability
to undergo a disk instability episode.  We further discuss the origin
and evolution of disk instabilities in the next section.

\subsection{Disk instability in central and satellite galaxies}
\label{sec:history_DI}

In the previous sections, we have found that the contribution of disk
instability to bulge growth is significant for galaxies with
$0.5<B/T<0.7$ and $M_*\in[10^{10.2};\;10^{10.8}]\;{\rm M_{\sun}}$.
Fig.~\ref{fig:history_BT_med} and Table~\ref{tab:DI_mer} show that the
contribution of this bulge formation channel is important also for
massive central galaxies with intermediate morphology ($0.5<B/T<0.7$).
As we have seen earlier, the relevance of the disk instability channel
reflects in the size--mass relation (because DI bulges are smaller than
merger bulges). The effect is more important at intermediate masses
because in this mass bin galaxies with $0.5<B/T<0.7$ are more numerous
than those with $B/T>0.7$. 

In Sec.~\ref{sec:r_ms_cen_sat}, we have shown that, for intermediate
mass ET galaxies, centrals selected using a $B/T=0.5$ threshold have
on average a smaller half-mass radius than the overall population of
ET galaxies.  This is in part due to the fact that bulges of central
galaxies have a slightly larger contribution from disk instability
with respect to those of the overall population ($\sim 50\%$ in
centrals and $\sim46\%$ in all galaxies).  Furthermore, bulges and
disks in central galaxies that underwent disk instabilities have
smaller sizes (by $\sim 0.7$ dex) than those formed in satellites of
the same mass and $B/T$. Below, we discuss the origin of these
differences in more detail.

\begin{figure*}
    \includegraphics[trim=1.1cm 0.3cm 1.2cm 0.9cm, clip, width = 0.99\columnwidth]{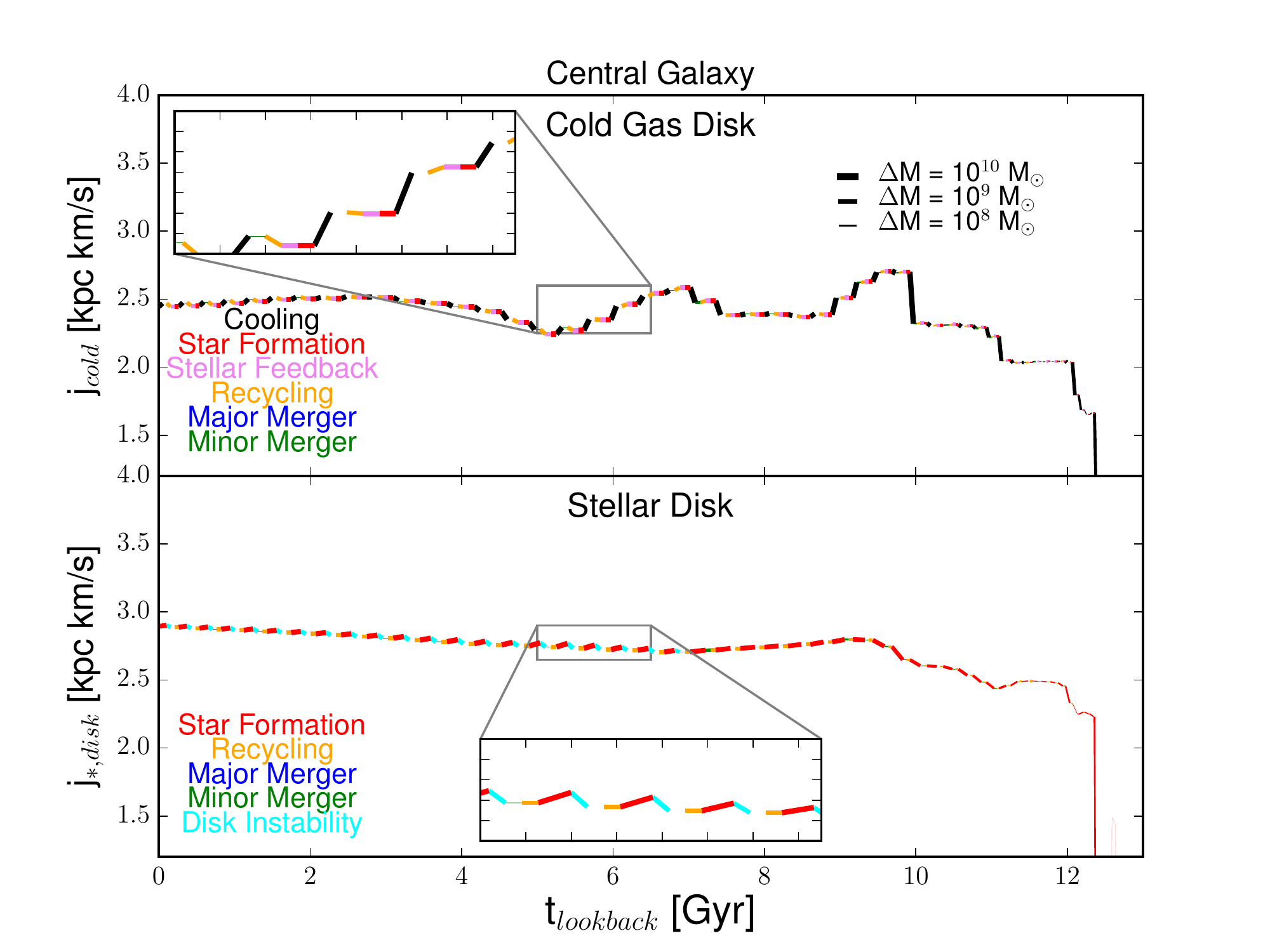}
    \includegraphics[trim=1.1cm 0.3cm 1.2cm 0.9cm, clip, width = 0.99\columnwidth]{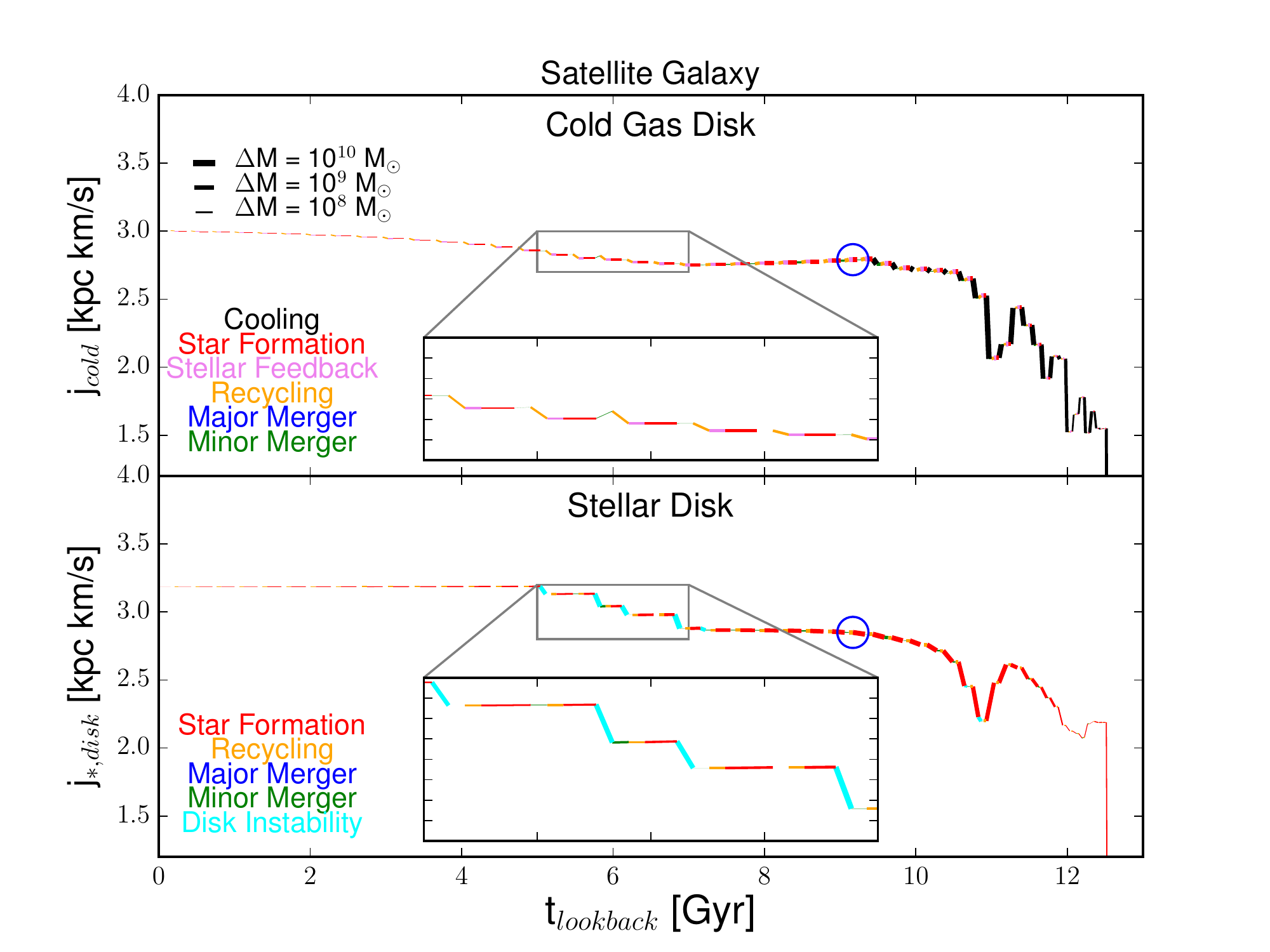}
    \caption{
      Evolution of the specific angular momentum of two representative
      galaxies with $10^{10.2}<M_*<10^{10.8}\;{\rm M_{\sun}}$ and
      $0.5<B/T<0.7$.  The left panels show the evolution of a central
      galaxy, while the right panels are for a satellite.  The top
      panels show the evolution of the specific angular momentum of
      the cold gas disk, while the bottom panels show that of the
      stellar disk.  The colored segments represent variations of the
      specific angular momentum, and are color coded according to the
      physical process causing them (see legend). The thickness of
      each segment increases with the absolute magnitude of the mass
      variation resulting from each process.  The zoomed-in regions
      highlight recursive disk instability events.}
    \label{fig:history_cen_sat}
\end{figure*}

Fig.~\ref{fig:history_cen_sat} shows the evolution as a function of
lookback time of the specific angular momentum of the gaseous disks
(top panels) and of the stellar disks (bottom panels) for a central
(left panels) and a satellite galaxy (right panels).  
These test 
galaxies have been selected within $10^{10.2}<M_*<10^{10.8}\;{\rm
  M_{\sun}}$ and $0.5<B/T<0.7$, and are representative of the whole sample.  
  Each segment shows a
variation of specific angular momentum due to a specific physical
process (different colors correspond to different processes, as
indicated in the legend). Let us focus first on the evolution of the
central galaxy (left panels).  At early times ($t_{\rm lookback}>10\;{\rm
  Gyr}$), the specific angular momentum of the gaseous disk grows due
to cooling (black lines in the top panel).  The specific angular
momentum of the cold gas is unaffected by star formation and stellar
feedback (red and magenta lines), and only slightly decreases due to
recycling (orange lines).  The angular momentum of the cold gas
follows the variation of the angular momentum of the parent halo: it
starts decreasing after $t_{\rm lookback}\sim10\;{\rm Gyr}$, increases
again at $t_{\rm lookback}\sim7\;{\rm Gyr}$, and then decreases again
until $t_{\rm lookback}\sim5\;{\rm Gyr}$. The stellar disk acquires the
angular momentum of the cold gas through star formation (bottom left
panel). When $j_{\rm *,disk}$ decreases, the disk contracts until it
eventually becomes unstable.  As described in
Sec.~\ref{sec:bulge_formation}, during disk instability events, part
of the disk mass is transferred to the bulge to restore the
stability. Since we assume angular momentum is conserved, this
increases the specific angular momentum (and size) of the disk. These
events are shown as cyan lines in the bottom left panel of
Fig.~\ref{fig:history_cen_sat}.  During subsequent star formation
episodes (see the zoom-in panel), the stellar disk mass increases
again, the cold gas decreases, and so does the specific angular
momentum. This triggers a new disk instability episode.  The
size of the galaxy therefore oscillates slightly, due to a series of
consecutive disk instability events.

In the case of the satellite galaxy (right panels), the early
evolution of the gaseous and stellar disks is similar to that of the
central galaxy.  After accretion (the last time the galaxy is a
central - marked as a blue circle in the figure), the angular momentum
of the cold gaseous disk cannot be affected by cooling, which is
suppressed. The stellar disk follows the evolution of the cold gas due
to star formation.  After accretion, star formation still occurs, but
at increasingly lower rates (thinner red segments), because gas is
consumed and not replenished via cooling. Also in this case, the
instability criterion is eventually met, and the satellite undergoes a
disk instability episode.  Similarly to the case of the central galaxy
examined above, the satellite enters a recursive cycle of star
formation and disk instability episodes.  In this case, the stellar
disk specific angular momentum (and thus scale radius) remains almost
constant due to star formation, but the newly formed stars trigger a
new instability event.  This leads to an increase of the specific
angular momentum of the stellar disk.  In satellites, $j_{\rm cold}$ is
not lowered by cooling, and keeps growing due to recycling from the
stellar disk.  This translates into higher $j_*$ also for stars formed
from the cold gas disk.  This sequence of events stops only when the
star formation becomes negligible, because the cold gas is nearly
exhausted.

The two examples analyzed are representative of the total population:
statistically, disk instabilities affect only slightly the size of
central galaxies, while they generally lead to a slight increase of
the size of satellite galaxies.

\subsection{Angular momentum and cold gas}
\label{sec:history_gas}

\begin{figure}
\centering
    \includegraphics[trim=0cm 3cm 1cm 4cm, clip, width = 0.9\columnwidth]{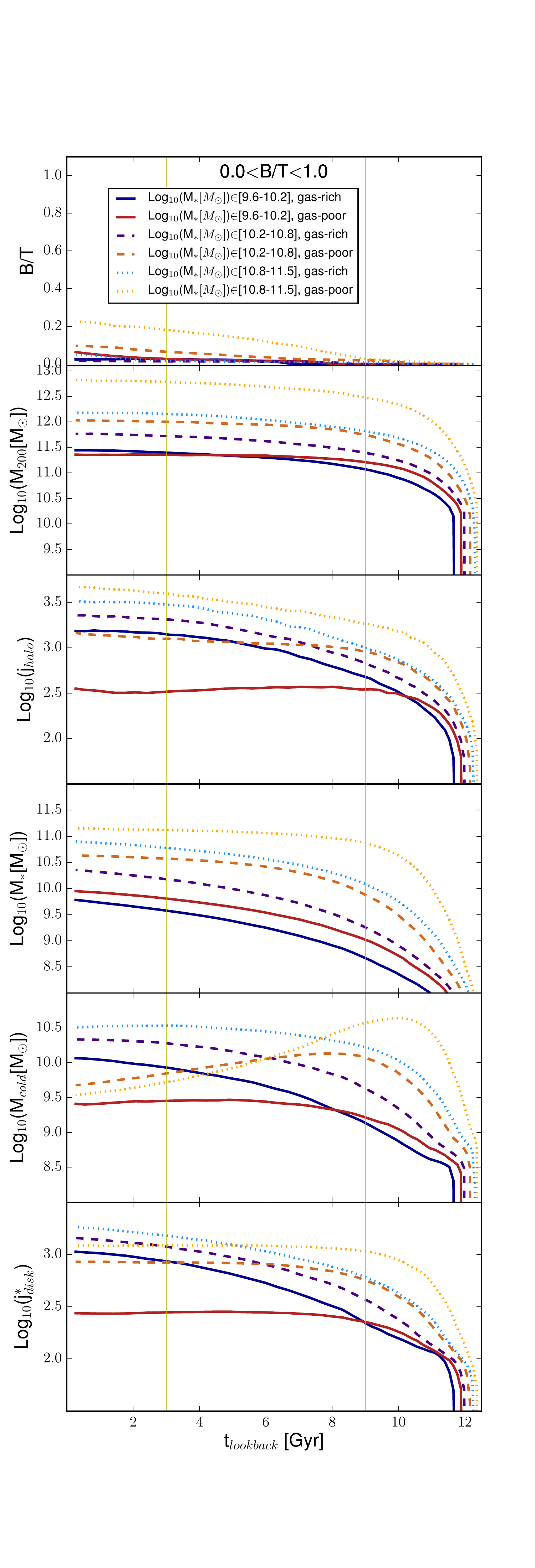}
    \caption{Median evolution as a function of lookback time of
      several galactic properties of model central galaxies in bins of
      stellar mass (different line-styles), and cold gas fraction
      (different colors).  From top to bottom: B/T, $M_{200}$,
      $j_{\rm halo}$, $M_*$, $M_{\rm cold}$ and $j_{\rm *,disk}$. Galaxies have
      been selected according to their stellar mass at redshift 0; gas
      poor/rich galaxies are selected as those below/above the
      16th/84th percentile of the distribution.}
    \label{fig:history_fgas}
\end{figure}

We have shown earlier that there is a strong dependence of the
specific angular momentum--mass relation on the cold gas fraction of
model galaxies. In this section, we investigate in detail the origin
of this dependence, by analyzing the evolution of gas-rich and
gas-poor central galaxies.  Also in this case we focus on central
galaxies, but we find consistent results for satellites. We show the
median evolution of some galaxy properties in
Fig.~\ref{fig:history_fgas}, as we did in
Fig.~\ref{fig:history_BT_med}.  We divide our model galaxies in the
same stellar mass bins at z=0: low ($M_*\in[10^{9.6}-10^{10.2}]\;{\rm
  M_{\sun}}$, solid), intermediate ($M_*\in[10^{10.2}-10^{10.8}]\;{\rm
  M_{\sun}}$, dashed), and high ($M_*\in[10^{10.8}-10^{11.5}]\;{\rm
  M_{\sun}}$, dotted lines). Results are qualitatively similar for
different values of $B/T$, so we average galaxies of different
morphological types. For completeness, we show the evolution for
different bins in B/T in Appendix~\ref{app:fcold}.  For each mass
sample considered, we have selected galaxies belonging to the extremes
of the $f_{\rm cold}$ distribution: i.e. we consider galaxies with
$f_{\rm cold}$ smaller than the 16$^{\rm th}$ percentile of the
distribution as gas-poor, and those with $f_{\rm cold}$ larger than the
84$^{\rm th}$ percentile as gas-rich.  Lines are color-coded according
to the stellar mass bin and the cold gas fraction, as indicated in the
legend.

For all mass bins considered, gas-poor galaxies are hosted by halos
that form earlier than those of gas-rich galaxies.  The halos hosting
gas-poor galaxies grow rapidly in mass, and acquire most of
their angular momentum during this phase of rapid accretion.  The
accretion history of halos hosting gas-poor galaxies translates into
large amounts of cold gas in these galaxies at early times, which
triggers significant early star formation.  Most of the stellar mass
of gas-poor galaxies is formed between 9 and 11 Gyrs ago.  At this
time, the specific angular momentum of the cold gas is relatively low,
like that of the parent dark matter haloes. In contrast, gas-rich galaxies
are hosted by halos that formed more recently than those hosting
gas-poor galaxies.  These halos accrete their mass more gradually,
and their mass increases down to very recent times.  As a consequence,
star formation occurs over a longer interval of time, and the stellar
disk can acquire the higher specific angular momentum of the halo at
late times.

\subsection{Dependence on physical prescriptions}
\label{sec:history_variants}
\begin{figure*}
  \centering
    \includegraphics[trim=2.5cm 0.1cm 3cm 0.5cm, clip, width = 0.9\linewidth]{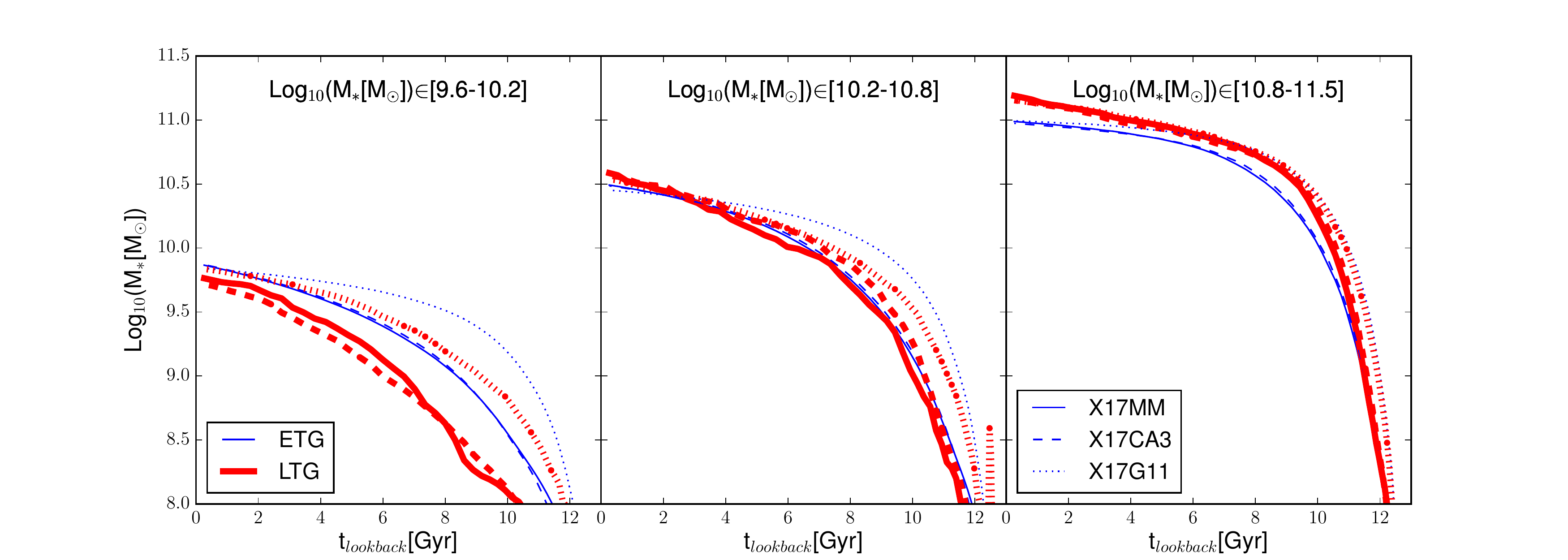}
    \caption{Median evolution of the total stellar mass as a function
      of lookback time , for galaxies in different stellar mass bins
      (different panels).  Galaxies are classified as LT (blue) and ET
      (red) using a threshold at $B/T=0.7$.  Different linestyles
      correspond to different models: solid for the X17MM model,
      dashed for X17CA3, and dotted for X17G11.}
    \label{fig:history_Ms_CA3_G11}
\end{figure*}

\begin{figure*}
\centering
    \includegraphics[trim=2.5cm 0.1cm 3cm 0.5cm, clip, width = 0.9\linewidth]{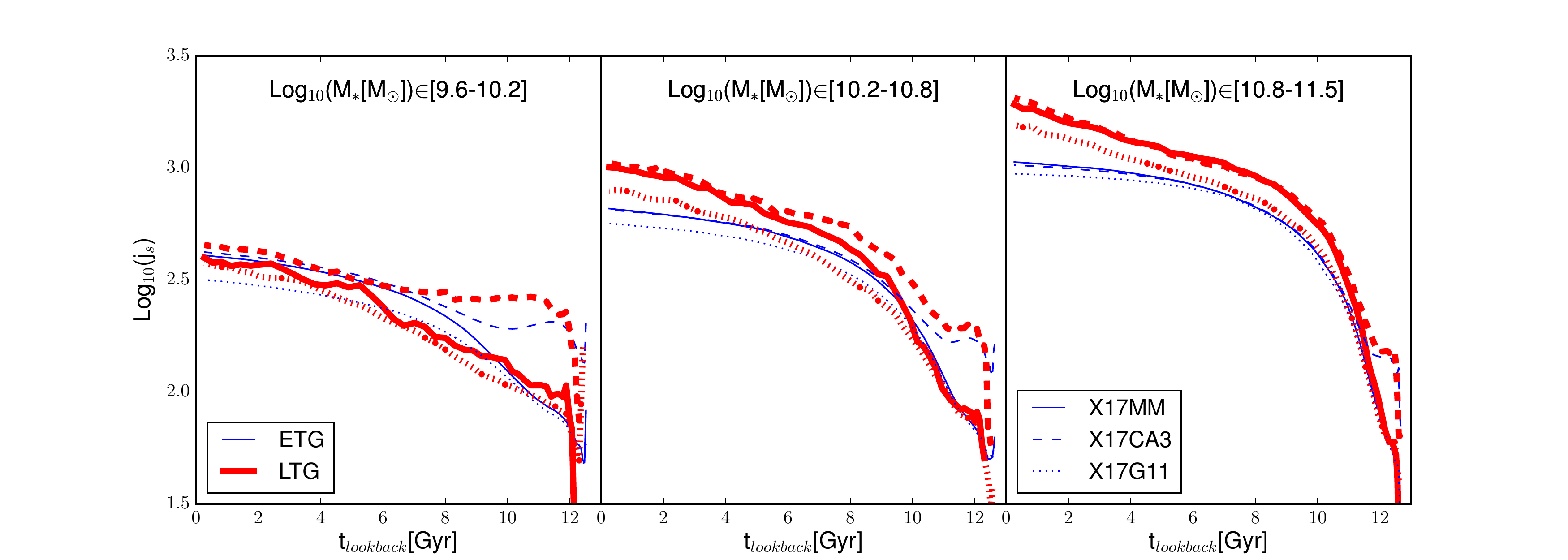}
    \caption{As in Fig.~\ref{fig:history_Ms_CA3_G11}, but this time
      showing the median evolution of the stellar disk specific
      angular momentum as a function of lookback time.}
    \label{fig:history_js_CA3_G11}
\end{figure*}

In this section, we study the origin of the different size--mass and
specific angular momentum--mass relations for the fiducial model X17MM
and for its variants X17CA3 and X17G11.  In
Fig.~\ref{fig:history_Ms_CA3_G11} and \ref{fig:history_js_CA3_G11}, we
show the evolution of the total stellar mass and of the stellar disk
specific angular momentum for ET (red) and LT (blue) galaxies, divided
using a $B/T=0.7$ cut.  Predictions from different models are shown
using different line styles (X17MM with solid, X17CA3 with dashed, and
X17G11 with dotted lines).

In the X17CA3 run, the gas cooling in the rapid mode regime has an
angular momentum three times larger than in the X17MM run. As we have
discussed earlier, this leads to slightly larger sizes and angular
momenta for low-mass galaxies ($M<10^{10.2}\;{\rm M_{\sun}}$).  When
considering the evolution of X17CA3 galaxies (dashed lines in the
figures), we see that their specific angular momentum is higher than
in the X17MM model during the first 2-3 Gyrs, as expected.  This,
however, does not imply a higher star formation rate, because a larger
$j_{\rm cold}$ translates into a larger disk radius and, as a consequence, a
lower gas surface density.  As a result, the stellar mass of galaxies
in the X17CA3 run evolves as in X17MM, although the specific angular
momentum of the stellar disk is much higher at early epochs ($j_*$
follows $j_{\rm cold}$ through star formation).  After 2-3 Gyrs, cold
accretion is no longer the dominating accretion mode, and the specific
angular momentum of the cold gas converges to approximately the same
values found in the X17MM run.

For the X17G11 run, we found a lower final specific angular momentum
for LT galaxies and, as a consequence, a lower normalization of the
size--mass relation.  In Fig.~\ref{fig:history_Ms_CA3_G11}, we find
that X17G11 galaxies form the bulk of their stars at earlier times
compared to their counter-parts in the X17MM run, for all the mass and
morphology bins considered.  The feedback scheme adopted in the X17G11
run allows ejected gas to be reaccreted earlier than in X17MM,
resulting in larger amounts of gas cooling onto the model galaxies in
the first 1-2 Gyrs.  This translates into significant early star
formation. After the initial peak, star formation gradually decreases,
while in the fiducial model it remains almost constant down to the
present day.  This is because in the X17MM run, ejected gas is
reaccreted more gradually.  Therefore, most of the stars in the disk
of X17G11 galaxies form from gas with lower specific angular momentum
than in the X17MM run, explaining the systematic offsets we find for
sizes and angular momenta of model LT galaxies. In
Sec.~\ref{sec:r_ms_CA3_G11}, we have shown that ET galaxies in the
X17G11 run have, on average, larger sizes than those of the same mass
in the X17MM run.  This is contrary to what one would expect
considering the evolution of the specific angular momentum just
discussed. For ET galaxies, however, one has to consider also the
different merger and disk instability histories in the two runs. We
find that in the X17G11 run, the star formation peak occurs earlier
than for X17MM, and this tends to decrease the occurrence of later
disk instability episodes. As a consequence, more
bulges in the X17G11 run are formed mainly during mergers with respect
to X17MM, which leads to an average bulge size in the X17G11 run
larger than in X17MM. When evaluating the specific angular momentum of
ET galaxies in the X17G11 model, the larger bulge sizes affect the
integration, because, at fixed mass, the bulge profile is flatter.
This enhances the
importance of rotational velocity at large radii in the calculation of
$j_*$. The rotational velocity increases with radius, and the
integrated final specific angular momentum is thus higher than in the
X17MM model.

\section{Discussion and conclusions}
\label{sec:conclusions}
In this work, we have analyzed the structural and dynamical properties of
galaxies in the framework of a state-of-the-art semi-analytic model.  In
particular, we have used the latest version of the GAlaxy Evolution and
Assembly \citep[GAEA][]{hirschmann2015} model. This includes a sophisticated
treatment for the non-instantaneous recycling of gas, metals, and energy
\citep{delucia2014}, and a new stellar feedback scheme partly based on results
from hydrodynamical simulations \citep{hirschmann2015}.  The model we have used
also includes a treatment for the atomic to molecular gas transition based on
the \citet{blitz2006} empirical relation, and a molecular hydrogen based star
formation law \citep{xie2017sam}.  Furthermore, our model includes
prescriptions to follow the angular momentum exchanges among galactic
components, and evaluates disk sizes from specific angular momenta, as in
\citet{guo10}.  In the following subsections, we discuss our results focusing
on: (i) comparison between model predictions and the observed size--mass and
angular momentum--mass relations; (ii) relevance of disk instability and limits
of our modeling approach; and (iii) dependence on different physical
prescriptions.

\subsection{Size and specific angular momentum}

Our galaxy formation model traces explicitly the exchanges of specific
angular momentum between different galactic components, and this
information is used to evaluate the disk scale radius. Therefore, the
size--mass and the angular momentum--mass relation of model disk
galaxies are strictly related to each other. For bulge dominated
galaxies, whose sizes are estimated using energy conservation arguments
and for which the contribution from disk instability can be important,
the correlation between size and angular momentum is less trivial,
also due to the approximations necessary to estimate the latter
quantity (see Appendix~\ref{app:j_estimates}).

Previous studies focused on the size--mass relation at relatively high
masses ($M_{*}>10^{10}\;{\rm M_{\sun}}$), and highlighted the
necessity of a specific treatment for gas dissipation during mergers
to obtain realistic bulge sizes \citep{shankar2014,tonini2016}.  In
our study, we have extended the comparison with observational results
down to stellar masses of $\sim 10^{9}\;{\rm M_{\sun}}$. For late type
galaxies, our predicted size--mass relation is in fairly good agreement
with recent observational estimates. In agreement with previous
studies, we find that dissipation during mergers is necessary to
correctly reproduce the size of early type galaxies, especially at low
stellar masses ($M_*<10^{10}\;{\rm M_{\sun}}$).  In our model, we find
that dissipation must be limited to major mergers, otherwise the
predicted bulge sizes are too small compared to observational
measurements. This assumption is reasonable since the prescriptions we
have used are based on binary merger simulations, with relatively
large mass ratios.  These simulations show that the effect of
dissipation is more relevant in major mergers between spiral galaxies,
and influences only slightly minor mergers between spiral galaxies
\citep[see][ and their Table~1, for a summary of their
  results]{porter2014dissipation}.  Our results are also in agreement
with those by \citet{shankar2014}, who applied the same treatment for
dissipation, only during major mergers, to the semi-analytic model by
\citet{guo10}.

In the framework of our model, the specific criteria used to select
late and early type galaxies affect significantly the predicted
size--mass relation, which appears to be in contrast with observational
findings. Specifically, we find that using a selection based on the
bulge-to-total mass ratio (B/T) leads to two well separated relations
for late and early type galaxies, and to a reasonable agreement with
observational measurements. In contrast, when selecting galaxies on
the basis of their specific star formation rate (sSFR), model early
type galaxies are offset high with respect to data, and their
size--mass relation does not differ significantly from that obtained
for active model galaxies. 
If we consider only central galaxies, the size--mass relation predicted
adopting a sSFR selection for early type galaxies is in agreement with
results based on a $B/T$ selection.  The treatment of satellites in
our model is rather simple: after the accretion, the hot gas reservoir
is instantaneously stripped and assigned to the central halo,
suppressing cold gas refueling through cooling. This treatment
suppresses the star formation in satellite galaxies, but does not
affect their morphology and size. Our model does not include a
treatment for physical processes such as tidal
interactions, which can effectively remove material from the galaxy
outskirts.  The introduction of these processes could reduce the sizes
of satellite galaxies and even affect their morphology.

Our model predictions for the specific angular momentum versus mass relation
agree fairly well with observational measurements in terms of slope, but are
offset low with respect to data both for early and late type galaxies. For
bulge dominated galaxies, this offset can be partially explained by the fact
that we assume our model bulges do not rotate. For late type galaxies this
could be explained, at least in part, by selection biases towards gas-rich
galaxies.  In fact, we find a relatively strong correlation between the
specific angular momentum and the gas fraction.  This correlation is a
by-product of halo and galaxy evolution: gas poor galaxies form most of their
stars at high redshift where the angular momentum of the parent halo is low,
while gas rich galaxies tend to form stars over a longer time-scale allowing
larger values of angular momentum.  We found a similar correlation in
\citet{zoldan2017MNRAS.465.2236Z}, where we have shown that the neutral atomic
hydrogen content of our model galaxies is correlated with the halo spin
parameter.  These results highlight how the halo evolution influences at the
same time the specific angular momentum and cold gas evolution.  Future
surveys, with a higher completeness and better controlled selection, will
provide us with more robust estimates of the specific angular momentum versus
mass relations. These results will allow us to identify the origin of the
discrepancies discussed above.

 The success of our model in reproducing the specific angular momenta of
  galaxies is rather impressive, given the intrinsic limitations of the
  semi-analytic approach in treating spatially resolved quantities. The same
  subject has been addressed in the framework of only a few other models.
  \citet{lagos2015sam_j} analyzed the misalignment between the specific angular
  momenta of the cold gas, of the stellar disks, and of the DM halo.  In their
  work, they used a modelling for angular momentum exchanges similar to ours,
  but added a specific post-processing treatment for spin flips during galaxy
  mergers/accretions.  They did not compare, however, the amplitude of the
  predicted specific angular momenta with observational estimates.
  \citet{stevens2016darksage} used a more sophisticated modelling for the
  evolution of the disk specific angular momentum. Specifically, they divided
  the galactic disk in annuli of different specific angular momentum, and
  evolved them according to physical processes applied to the individual annuli.
  Their predicted $j_{{\rm *,disk}}$--$M_*$ relation for spiral galaxies agrees
  well with observational measurements by \citet{fall2013js} and
  \citet{obreschkow2014jmbt}.  They found that the normalization of the relation
  is strongly affected by disk instabilities: the specific angular momentum
  decreases when disk instability is turned off. Our disk instability treatment
  appears to have a similar effect on $j_{{\rm *,disk}}$, but its influence on
  the evolution of LT galaxies is overall less important.
  \citet{stevens2016darksage} did not model the specific angular momenta of
  galaxies with $B/T>0.3$ so that their results cannot be directly compared to
  ours.

Our comparison with observational data has been limited to the local
Universe. We plan to extend this comparison to higher redshift in future
work. We note, however, that the strong correlation between the specific
angular momentum of the stellar disk with that of the DM halo suggests an
evolution similar to that observed, at least for late type galaxies.

\subsection{Disk instability and bulge size}
\label{sec:discussion_di}

In our model, central bulge-dominated galaxies have unrealistically small sizes
in the stellar mass range $10^{10}-10^{10.8}\;M_{\sun}$.  This is not the case
when considering the entire early-type galaxy population (i.e. including
satellite galaxies), and is due to different contributions and sizes of bulges
that form predominantly through disk instability or mergers. Specifically, we
find that most of the intermediate bulges ($0.5<B/T<0.7$) form mainly through
disk instabilities, while the largest ones ($B/T>0.7$) form mainly through
(major) mergers. The sizes of our model bulges formed through mergers are in
nice agreement with observational estimates (with a relatively small
  under-estimation with respect to observational data at large stellar masses), while disk
instabilities produce systematically smaller bulges, especially for central
galaxies. We find that disk instability events are typically associated with
haloes that suffer, at some point of their life-time, a decrease of specific
angular momentum.  As the hot gas is assumed to have the same specific angular
momentum as the dark matter halo, cooling transfers this loss of specific
angular momentum to the cold gas disk, and star formation transfers it to the
stellar disk. This eventually triggers recursive events of disk instability.
In our satellite galaxies, disk instabilities tend to result in a net increase
of stellar disk sizes, because cooling is suppressed after accretion. For
central galaxies, the size of the stellar disk component oscillates slightly
but is not significantly different from the value it had before the instability
episodes.

The unrealistic sizes obtained for bulges formed through disk
instability highlight the need to revise the prescriptions adopted for
this particular physical process.  The instability criterion we
currently use is based on old two-dimensional N-body simulations of a
purely stellar disk in a rigid halo \citep{efstathiou1982DI}.
Although widely adopted in the semi-analytic framework, this criterion
is not consistent with results from more recent N-body simulations
\citep[][and references therein]{athanassoula2008DI}.  At present,
however, no alternative prescription is available.

The bulge size estimation during disk instabilities should also be
revised. Secular processes are believed to give origin to thick, disk-like,
rotating `pseudo-bulges'.  These differ from classical, merger originated
bulges in their dynamical properties and in their stellar population
\citep[see][for a review]{kormendy2004PB}.  The scale length of classical and
pseudo-bulges are, however, comparable
\citep{fisher2008PB,gadotti2009_sdss_pseudo}, contrary to predictions from our
model. Recently, \citet{tonini2016} considered an explicit division between
classical and pseudo bulges in their semi-analytic model, assuming disk-like
structural properties for pseudo-bulges.  In their model, pseudo-bulges have
masses similar to those of classical bulges, but are concentrated in the
$M_B\in[10^{10},10^{11}]\;{\rm M_{\sun}}$ range.  In general, their
pseudo-bulges have small half-mass radii (up to $5$ kpc), while classical
bulges are larger (up to $20$ kpc).  In our model, disk instability bulges are
much smaller than 1 kpc.  Assuming that stars in disk instability bulges are
distributed according to an exponential profile, we obtain half-mass radii
larger than those represented in Fig.~\ref{fig:reff_components} of about $\sim
0.5$ dex.  A consistent treatment of a two components bulge would allow the
formation of larger bulges, because the larger radii of disk instability bulges
would enter the energy equation during mergers. Furthermore, a rotating
pseudo-bulge would allow for an explicit treatment of fast and slow rotators,
and model results could be consistently compared with results from recent
integral field spectroscopic surveys, as ATLAS$^{\rm 3D}$ and SAMI. 
  Similar studies have been carried out in the framework of semi-analytic
  models, but without an explicit analysis of the specific angular momentum of
  fast and slow rotators \citep[see e.g.][]{khochfar2011FR_SR}.  We postpone a
self-consistent implementation of a two-bulge model to a future work.

\subsection{Dependence on physical modeling: stellar feedback and cooling} 

Our analysis confirms that the dynamical properties of galaxies depend
strongly on the galaxy star formation and assembly history.  We have
tested the relative importance of specific prescriptions regulating
the baryon cycle. Specifically, we have considered an alternative
stellar feedback scheme, and the influence of a higher specific
angular momentum for gas cooling in rapid mode.

In the alternative feedback scheme considered, the ejected gas is
re-accreted earlier than in our fiducial model.  This causes the bulk
of star formation to occur earlier than in our reference model. Since
the halo angular momentum is generally lower at higher redshift, the
alternative scheme translates into a lower specific angular momentum.
Several studies based on hydrodynamic simulations have highlighted
that strong stellar feedback at early times leads to high final
specific angular momentum of the stellar disk
\citep{ubler2014j_fs,hirschmann2013}.  In these simulations, the gas
ejected through stellar winds is accelerated and re-accreted with
angular momentum higher than it had when it was ejected. Our model
does not include such a sophisticated treatment for ejected gas.
Simply, the reaccreted gas has the same specific angular momentum of
the dark matter halo at the time of gas re-accretion. This means that
the re-accreted gas has indeed an angular momentum typically larger
than at the time of ejection. Our results confirm the fundamental role
of stellar feedback in determining the dynamical properties of
galaxies, by regulating the time when most of the stars are formed
(and therefore most of the angular momentum is acquired).

We have also analyzed the influence of a higher specific angular
momentum for gas cooled in the rapid cooling regime.  This was
motivated by results from recent hydrodynamic simulations
\citep{stewart2011cold_accr,pichon2011cold_accr,danovich2015cold_accr},
that have highlighted that this gas can have angular momentum from 2
to 4 times larger than that of the parent dark matter halo
halo. Including this modification in our model, the star formation
history of model galaxies is not significantly affected.  We find that
small galaxies are those mostly affected, with slightly larger sizes
and angular momenta than those in our reference model.  Therefore, we
expect this physical process to be important at higher redshift, where
a larger fraction of the population is dominated by the cold accretion
mode.

\subsection{Summary}

We have shown that results from our GAlaxy Evolution and Assembly
(GAEA) model, modified to include a treatment for gas dissipation
during major mergers, are in quite nice agreement with the observed
size--mass and specific angular momentum--mass relation observed in the local
Universe. The main conclusions of our work can be summarized as follows:

\begin{itemize}
\item The predicted size--mass and specific angular momentum--mass
  relations are in fairly good agreement with observational
  measurements when late and early type galaxies are selected using a
  cut in bulge-to-total (B/T) stellar mass. A late/early type selection
  based on the specific star formation rate (sSFR) would instead
  include, in our model, too many quenched disky satellite galaxies,
  leading to a too high normalization of the size--mass relation for
  early type galaxies.
\item The sizes of our model bulges are strongly affected by their
  dominant formation channel. Bulges formed mainly through disk
  instabilities have unrealistically small sizes. This affects, in
  particular, central galaxies with $0.5<B/T<0.7$ and
  $M_*\in[10^{10};\;10^{10.8}]\;M_{\sun}$. We have discussed how an
  explicit treatment for a two-bulge component could lead to a better
  agreement with data.
\item The stellar specific angular momentum of a galaxy is strongly
  correlated with its cold gas fraction.  This correlation originates
  naturally from the galaxy accretion history: in gas-poor galaxies
  today, most of the stars are formed at early times when the halo
  angular momentum is low. In contrast, gas-rich galaxies form their
  stars over a more extended time-scale, which allows larger angular
  momentum to be incorporated in the stellar component of galaxies.
\item The adopted stellar feedback scheme can affect significantly the
  galaxy star formation history, and therefore the predicted angular
  momenta and sizes of model galaxies. A shift to higher (lower)
  redshift of the peak of star formation leads to a lower (higher)
  normalization of the angular momentum/size--mass relation for late
  time galaxies. For early-type galaxies, the shift might be less
  important or even go in the opposite direction, in our model, due to
  a different contribution from the disk instability channel.
\item A different initial specific angular momentum of cold gas
  accreted through rapid cooling regime influences the sized and
  angular momenum of small galaxies today, and is expected to have an
  important impact at higher redshift where a larger fraction of the
  galaxy population is dominated by this accretion mode.
\end{itemize}


\section*{Acknowledgements} 
We thank Barbara Catinella, Luca Cortese and Claudia Lagos for useful and
interesting discussions.  GDL and LX acknowledge financial support from the
MERAC foundation.



\bibliographystyle{mnras}
\bibliography{biblio_1} 


%
\appendix
\section[Profiles]{Mass and velocity profiles}
\label{app:profiles}

We assume that the stellar and gaseous disks are described by
exponential surface density profiles:
\begin{equation}
\label{eq:m_disk}
 \Sigma_{\rm disk}(r) = \Sigma_0 e^{-\frac{r}{R_{\rm disk}}}
\end{equation}
where $R_{\rm disk}$ is the scale radius, and $\Sigma_0=M_{\rm *,disk}/(2\pi
R_{\rm disk}^2)$ the central surface density. For the bulge, we assume a
stellar distribution that follows a Jaffe profile
\citep{jaffe1983bulge}:
\begin{equation}
\label{eq:m_bulge}
 \rho(r)=\frac{M_B}{4\pi R_B^3}\left(\frac{r}{R_{B}}\right)^{-2}\left(1+\frac{r}{R_{B}}\right)^{-2}
\end{equation}
where $R_{B}$ is the scale radius, and $M_B$ is the mass of the bulge.
The half-mass radii ($R_{1/2}$) of model galaxies are obtained
projecting the stellar mass profiles in Eq.s~\ref{eq:m_disk} and
\ref{eq:m_bulge} assuming galaxies are seen face on, and calculating
the radius that encloses half of the projected mass.

To estimate the galaxy specific angular momentum, we assume that the
bulge is dispersion supported ($v_{\rm bulge}(\vec{r})=0$), and that the
stellar disk is symmetric and supported by rotation:
\begin{equation}
\label{eq:vel}
 v_{\rm rot}^{\rm disk}(r) = \sqrt{\frac{GM(<r)}{r}},
\end{equation}
where $G$ is the gravitational constant, and $M(<r)$ is the total mass
enclosed within $r$.  It includes the stars in the disk and in the
bulge, the gas in the disk, and the corresponding fraction of the
parent dark matter halo.  For the latter component, we assume a
Navarro-Frenk-White profile \citep{NFW1996}:
\begin{equation}
 \label{eq:NFW}
 \rho(r) = \frac{\rho_0}{\frac{r}{R_{\rm DM}}\left( 1+\frac{r}{R_{\rm DM}} \right)^2},
\end{equation}
where $\rho_0$ and $R_{\rm DM}$ are a density parameter and the scale
radius of the halo. Both can be estimated using the concentration
parameter of the halo, which we calculate using the correlation between
$M_{200}$ and concentration published in
\citet{neto2007DMconcentration}. This relation has a large scatter,
but we checked that our results are not significantly affected by this
using the extremes of the distribution, instead of the median.  The
virial radius is directly proportional to the scale radius:
$R_{200}=cR_{\rm DM}$, while the density parameter $\rho_0$ can be obtained
integrating Eq.~\ref{eq:NFW} to $R_{200}$ and forcing it to be equal
to $M_{200}$. The latter quantity is measured directly from the
simulations for all haloes hosting central galaxies, while for
satellites it corresponds to the particle mass times the number of
bound particles in the parent subhalo.  $R_{200}$ is calculated from
$M_{200}$, given the redshift of the halo.

\section{Specific angular momentum estimates}
\label{app:j_estimates}
For a comparison with observational measurements of the specific
angular momentum, we consider different estimates for our model
galaxies. We refer to the cartoons in Fig.~\ref{fig:j_cartoon}, to
help the reader in visualizing the description below.

The first estimate we consider is based on a three dimensional (3D)
model of the galaxy (see the top panel of Fig.~\ref{fig:j_cartoon}).
Using the 3D mass and velocity distributions for each galaxy, we
calculate the specific angular momentum within the radius $R$, by
integrating the velocity and mass profiles over all radii $r<R$.  The
specific angular momentum of the disk, obtained integrating on the
plane of the disk, is:
\begin{equation}
 j_{\rm disk}^{3D}(R) = \frac{\int_0^R r \; v_{\rm rot}^{\rm disk}(r) \;
   \Sigma_{\rm disk}(r)\; r\; dr}{\int_0^R \Sigma_{\rm disk}(r) \; r
   \; dr}
\end{equation}
If we include the bulge, for which we assume $v_{\rm bulge}(r)=0$, 
we obtain: 
\begin{equation}
 j_{\rm tot}^{3D}(R) = \frac{\int_0^R r \; v_{\rm rot}^{\rm disk}(r) \;
   \Sigma_{\rm disk}(r) r\;dr}{\int_0^R \Sigma_{\rm disk}(r) r \;dr + \int_0^R
   \rho_{\rm bulge}(r) 4\pi r^2 dr}
\end{equation}
where the integration for the bulge is carried out in 3D space. 

In observations, the 3D information is not available: the galaxy is
projected on the sky (2D), with a random inclination, and the velocity
information is typically available only along the line of sight
(l.o.s.).  In addition, it is difficult to separate the contributions
from the bulge and the disk.  We mimic this situation assuming all
model galaxies are edge-on (see the bottom panel of
Fig.~\ref{fig:j_cartoon}). 
In the 2D projection, we adopt a cartesian coordinate system centered at the galactic center, and aligned with the disk plane ($r$ coordinate) and the disk rotation axis ($z$ coordinate).
In this way, the l.o.s. velocity measured along the disk, at a projected distance $r$ from the center, is
exactly the rotational velocity at the 3D distance $r^{3D}=r$: $v_{\rm los}(r) =
v_{\rm rot}^{\rm disk}(r)$.  We project the stellar profiles of the stellar
disk and of the bulge on the edge-on plane, and sum them into a single
stellar component: $\Sigma_*(r,z) = \Sigma_{\rm disk}^{\rm edge-on}(r,z) +
\Sigma_{\rm bulge}(r,z)$. 
We also assume $h_{\rm disk} = R_{\rm disk}/7.3$ \citep{kregel2002hs}, and:
$$ v_{\rm los}(r,z) = \begin{cases}
               v_{\rm rot}^{\rm disk}(r) & {\rm if\;} z<h_{\rm disk}\\
               0	& {\rm if\;} z>h_{\rm disk}
              \end{cases} $$
In this way, the bulge fraction contained in a cylinder of height $h_{\rm
  disk}$ cannot be disentangled from the projected disk and, in the
  integration, it is assumed to rotate with the disk. We assume there is no
rotation outside the cylinder.  The specific angular momentum calculated along
a slit of the same height of the disk $h_{\rm slit} = h_{\rm disk}$ is:
\begin{equation}
 j_{\rm slit}^{2D}(R) = \frac{\int_0^R \int_0^{h_{\rm slit}} r \;
   v_{\rm los}(r,z)\; \Sigma_*(r,z)\;dr\;dz}{\int_0^R
   \int_0^{h_{\rm slit}}\Sigma_*(r,z)\;dr\;dz}
\end{equation}
Including also the bulge component outside the slit:
\begin{equation}
 j_{\rm tot}^{2D}(R) = \frac{\int_0^R \int_0^R r \;
   v_{\rm los}(r,z)\; \Sigma_*(r,z)\;dr\;dz}{\int_0^R \int_0^R\Sigma_*(r,z)\;dr\;
   dz}
\end{equation}
These estimates mimic the integrations performed for observed projected
galaxies, with $j_{\rm tot}^{2D}$ similar to an integration on circular or
elliptical concentric annuli, and $j_{\rm slit}^{2D}$ similar to an
integration along the major axis.
In the case of LT galaxies, a precise estimate of the galaxy inclination is usually possible, and the integration is performed using de-projected quantities. 
In the case of ET galaxies or spheroids, this is more difficult, as the precise measure of their rotational velocity. 
In this case a rough estimate of $j_*$ is given by the projected integration.

\citet{romanowsky2012js} found that the total specific angular momentum is well
approximated, after de-projection for inclination, by an empirical
formula that depends on the effective radius, the velocity measured at
two effective radii, and a factor $k_n$ that depends on the S\'ersic
index. In the case of a disk+bulge galaxy, they sum the contributions
from the disk and the bulge, weighting them for the corresponding
light (mass) fraction, $D/T$ and $B/T$:
\begin{equation}
\label{eq:j_RF_DB}
 j_{D+B} = k_{n_D} v_{\rm disk}(2R_{D}^e) R_{D}^e \frac{D}{T} + k_{n_B} v_{\rm bulge}(2R_{B}^e) R_{B}^e \frac{B}{T}
\end{equation}
In the above equation, $n_x$ and $R_x^e$ are the S\'ersic index and
the effective radius of the disk ($x=D$), or of the bulge ($x=B$).
The disk velocity $v_{\rm disk}(2R_{D}^e)$ is measured from the ionized
gas of the disk, while the bulge rotational velocity $
v_{\rm bulge}(2R_{B}^e)$ is estimated from its relation with the
ellipticity $\epsilon$ and central velocity dispersion $\sigma_0$,
through:
\begin{equation}
 v_{\rm bulge} = \left( \frac{v}{\sigma} \right) ^* \sigma_0 \left( \frac{\epsilon}{1-\epsilon} \right)^{1/2}
\end{equation}
$(v/\sigma)^*\sim0.7$ is a parameter describing the relative dynamical
importance of rotation and pressure, and a value equal to one
corresponds to that of an oblate isotropic system viewed edge-on
\citep{kormendy1982vsigma}.
The value of $0.7$ was chosen following the work by \citet{romanowsky2012js}.

We assume all model galaxies are composed of a disk+bulge, with a disk
S\'ersic index $n_D=1$ ($k_1=1.19$), $v_{\rm bulge}=0$, and $\epsilon=0$.
Thus the second part of Eq.~\ref{eq:j_RF_DB} is always zero and:
\begin{equation}
\label{eq:j_RF}
j^{RF} = k_1\; v_{\rm rot}^{\rm disk}(2R_D^e) \; R_D^e \frac{D}{T}
\end{equation}

The assumption of perfectly spherical, dispersion dominated bulges is very
strong.  We evaluate its impact by assuming, alternatively, that all bulges
have a S\'ersic index $n_B =4$ ($k_4=2.29$), an ellipticity $\epsilon = 0.2$
\citep[this is the median ellipticity of the elliptical/lenticular SDSS
  galaxies as found in ][]{hao2006ellipticity}, and a velocity dispersion
evaluated using the virial theorem:  $\sigma_0 = \sqrt{G(M_{B}+\delta
    M_{\rm disk})/(2R_{B})}$. In the last equation, $\delta M_{\rm disk}$ is
  the fraction of the disk that influences the bulge dynamics, which we assume
  to be the disk inside $R_{B}$ (higher values do not affect significantly
  model results).  Eq.~\ref{eq:j_RF_DB} then can be written as:
\begin{equation}
\label{eq:j_RF_epsilon}
 j^{RF}_{\epsilon = 0.2} = k_1 v_{\rm rot}^{\rm disk}(2R_D^e) R_D^e
 \frac{D}{T} + k_4 0.35 \sqrt{\frac{G(M_{B}+\delta M_{\rm disk})}{2R_{B}}} R_{B}^e
 \frac{B}{T}
\end{equation}

The different estimates computed can also be compared to the standard
output of our model ($j_{\rm disk}^{\rm SAM}$), weighted for the bulge
contribution: $j_{\rm tot}^{\rm SAM} = j_{\rm disk}^{\rm SAM}(1-B/T)$.

\begin{figure}
\centering
    \includegraphics[trim=0.1cm 0cm 1cm 0cm, clip, width = 0.9\columnwidth]{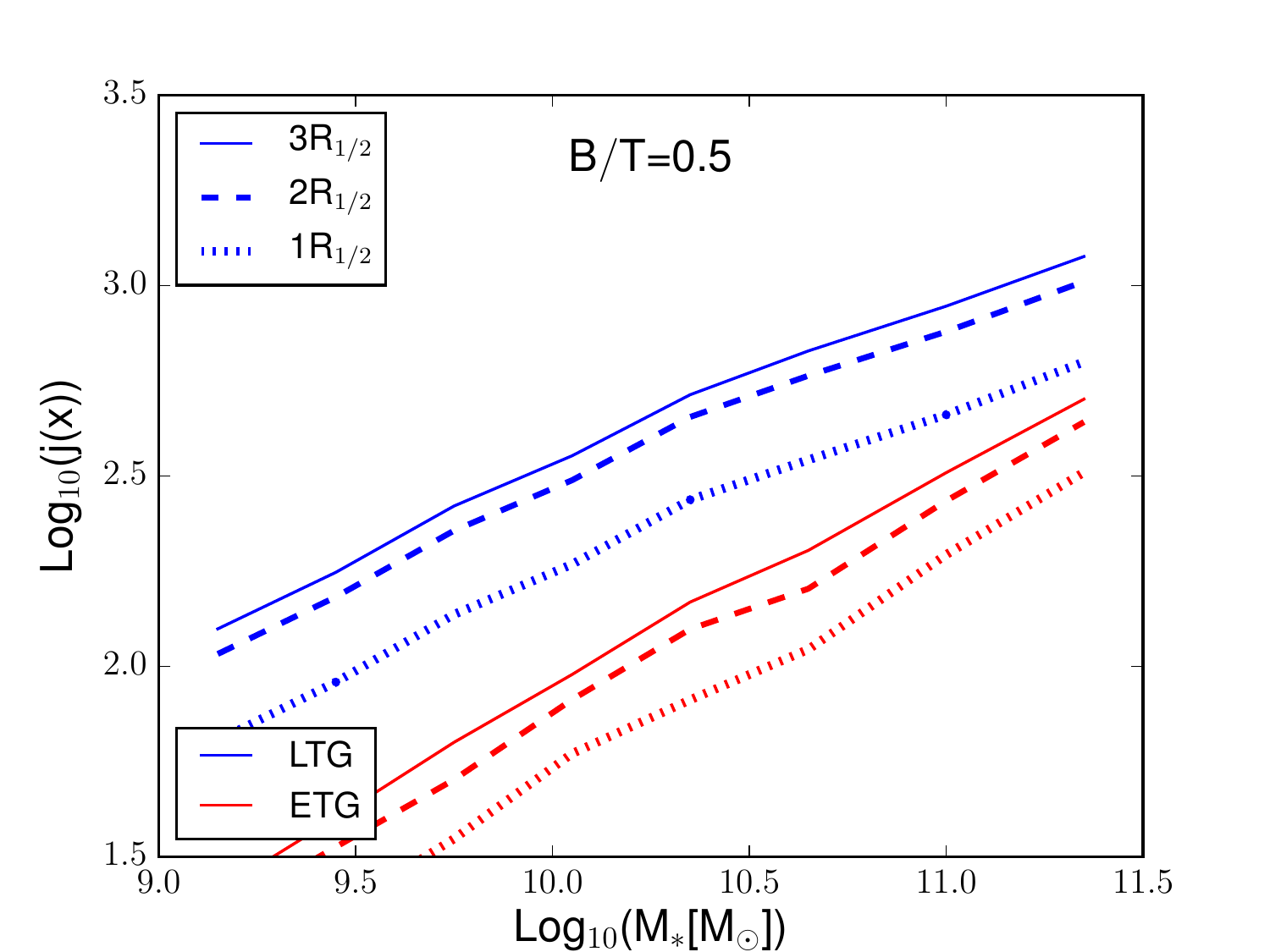}
    \caption{The $j_{\rm slit}^{2D}$--$M_*$ relation for LT and ET galaxies
      (blue and red), selected by their morphology ($B/T=0.5$), from
      the X17MM model.  The $j_{\rm slit}^{2D}(r)$ is evaluated
      considering projected profiles at different radii: $r= 1,2,3 \;
      R_{1/2}$.}
    \label{fig:j_ms_radii}
\end{figure}

Fig.~\ref{fig:j_ms_radii} shows the median $j^{2D}_{\rm slit}$--$M_*$
relation evaluated at 1, 2 and 3 $R_{1/2}$.  The difference between
2$R_{1/2}$ and 3$R_{1/2}$ is much smaller than that between 2$R_{1/2}$
and 1$R_{1/2}$.  In the paper, 
we show $j_*$ integrated out to a radius similar to that of the observational samples considered. 
When observations extrapolate to the total $j_*$, we use $j_*$ integrated out to $7\,R_{1/2}$, assuming its value has converged to the total value.
In the comparison between different runs of the model, we use 
the specific angular momentum corresponding to 2$R_{1/2}$, a distance
that provides a good compromise between convergence and typical observational
limits.

\begin{figure}
\centering
    \includegraphics[trim=0.1cm 0cm 1cm 0cm, clip, width = 0.9\columnwidth]{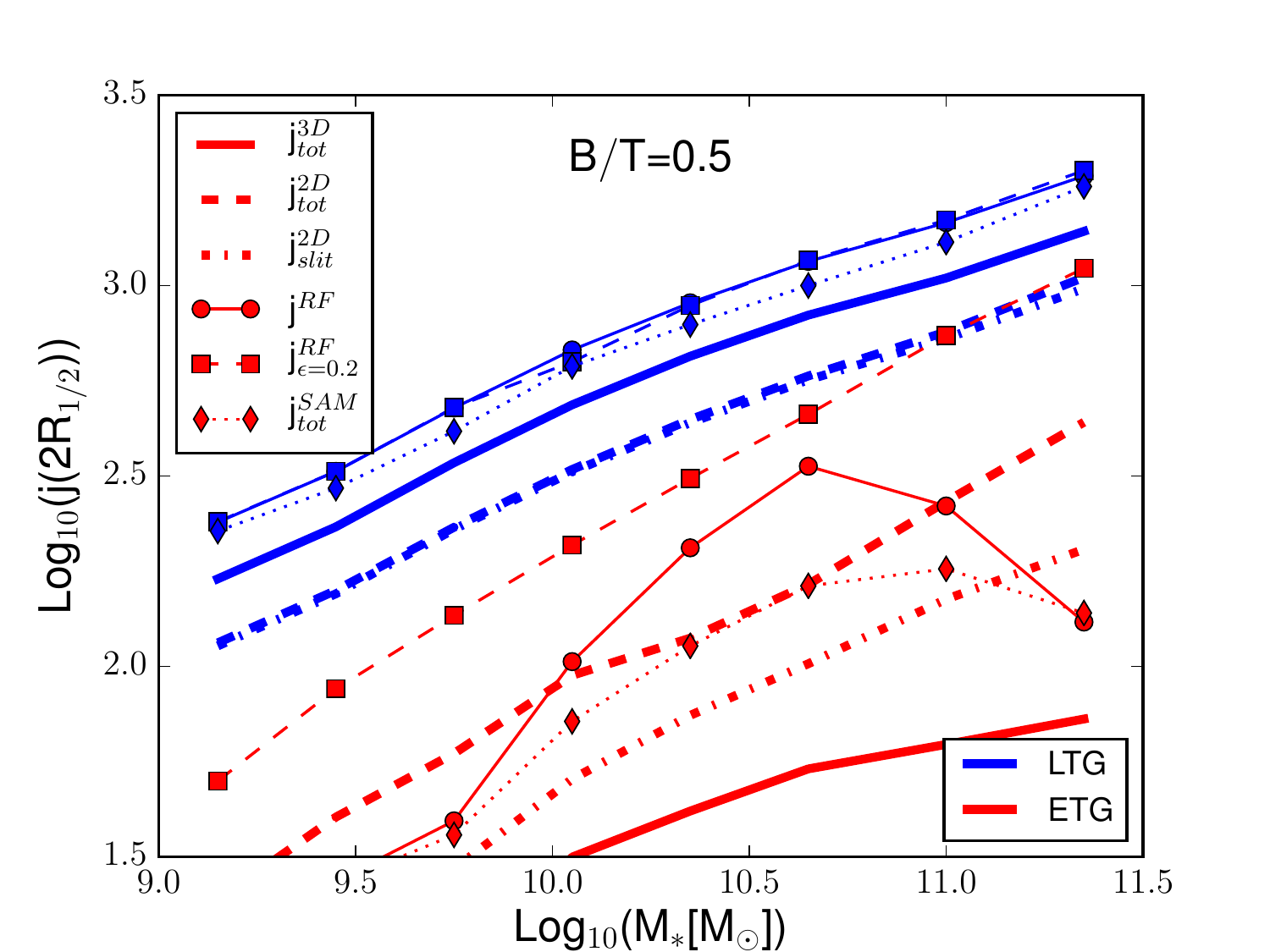}
    \caption{The median $j$--$M_*$ relation for LT and ET galaxies
      (blue and red) selected by their morphology ($B/T=0.5$).  All
      $j_*$ estimates are evaluated at $2R_{1/2}$.  Different lines
      and symbols represent different estimates for the specific
      angular momentum, as detailed in the text and as indicated in
      the legend.}
    \label{fig:j_ms_r_o_m}
\end{figure}
Fig.~\ref{fig:j_ms_r_o_m} shows the median $j$--$M_*$ relation
predicted for LT (blue) and ET (red) galaxies from the X17MM model.
We only consider a selection assuming a $B/T=0.5$ cut here, but we
have verified that results are qualitatively similar for alternative
selections.  Different line styles correspond to the different
estimates introduced earlier, as indicated in the legend.  All
relations obtained for LT galaxies, based on different estimates of
$j_*$, are above the ET relations. The estimates $j^{RF}$ and
$j^{RF}_{\epsilon = 0.2}$ give almost identical relations for LT
galaxies, and correspond to the highest normalization of the
$j_*$--$M_*$ relation for these galaxies.  This is expected, because
the influence of bulges in eq.s~\ref{eq:j_RF} and
\ref{eq:j_RF_epsilon} for galaxies with $B/T<0.5$ is negligible.  The
relation assuming the direct model output, $j^{\rm SAM}_{\rm tot}$, is only
slightly below that based on the empirical formula proposed by
\citet{romanowsky2012js}.  The relation based on the 3D estimate,
$j^{3D}_{\rm tot}$, is parallel to these but is offset low by $\sim 0.2$
dex.  Both the 2D estimates $j_{\rm slit}^{2D}$ and $j_{\rm tot}^{2D}$ lie on
the same relation, shifted 0.2 dex below that based on the 3D
estimate.  This is not surprising, because LT galaxies have a small
bulge, and a large fraction of it is contained in the slit.  The
difference with respect to the 3D estimate is due to the projection of
the disk mass: most of the disk mass, residing at the center, has a
lower velocity than in the de-projected case.  We expect that for
inclinations lower than edge-on this relation moves up towards the 3D
relation.  This argument is valid only for LT galaxies for which the
bulges, whose projected distribution is spherical and does not depend
on inclination, do not dominate the central stellar mass distribution.


While different estimates of $j_*$ for LT galaxies correspond to
parallel relations with relatively small shifts, estimates obtained
for ET galaxies cover a much larger region of the $j_*$--$M_*$ plane.  As
for LT galaxies, the highest normalizations are obtained for the
empirical estimates by \citet{romanowsky2012js},
$j^{RF}_{\epsilon=0.2}$ and $j^{RF}$.  The $j_*$--$M_*$ relation based on
the estimate that includes rotation in bulges,
$j^{RF}_{\epsilon=0.2}$, lies very close to the relation obtained for
LT galaxies employing the 2D estimates, with a slightly steeper slope.
\citet{romanowsky2012js} derived the formulae for $j^{RF}$ and $j^{RF}_{\epsilon=0.2}$ to describe disk+bulge galaxies. 
They analyzed $j_*$ of a small sub-sample of elliptical galaxies using alternative measurements, finding little convergence.
Nevertheless, they used the formula to estimate $j_*$ for their total sample. 
Therefore, the relation we show for ET galaxies should be intended as a rough estimate of our expectations for a realistic population of rotating bulges, with the caveat that the validity of the formula used has not been assessed for $B/T>0.5$.
As expected, the $j^{RF}_{\epsilon=0.2}$ estimate is larger than that
obtained assuming $\epsilon =0$ ($j^{RF}$).  The latter does not
include a bulge velocity component, and the calculation depends only
on the disk size and its velocity.  The dependence on the disk size is
important, because measuring the disk radius in a bulge dominated
galaxy is not easy.  In fact, using the effective radius of the galaxy
in the $j^{RF}$ estimations, instead of the disk effective radius,
lowers the relation at intermediate masses to the same position of the
relation based on the $j^{2D}_{\rm tot}$ estimate (for
$M_*<10^{10.7}\;{\rm M_{\sun}}$). The relation obtained considering
the direct model output for ET galaxies, $j^{\rm SAM}_{\rm tot}$, is close to
that obtained using $j^{RF}$.  The specific angular momentum increases
with increasing stellar mass up to $M_*\sim10^{10.7}\;{\rm M_{\sun}}$;
for larger stellar masses, the median value of $j^{\rm SAM}_{\rm tot}$ first
flattens and then decreases.  The relation assuming the 2D circular
estimate, $j_{\rm tot}^{2D}$, is parallel to that obtained using
$j^{2D}_{\rm slit}$, but is shifted up by 0.3-0.4 dex.  We tested the
influence of the slit height $h_{\rm slit}$ on the predicted $j_*$--$M_*$
relation, finding that a smaller $h_{\rm slit}$ would shift
$j_{\rm slit}^{2D}$ upwards, and $j_{\rm tot}^{2D}$ downwards.  For a slit
with height of $\sim 0.1h_{\rm slit}$, the shift of the relation using
$j_{\rm slit}^{2D}$ is of 0.1 dex, while that of the relation based on
$j_{\rm tot}^{2D}$ is of 0.2 dex - a modest effect.  The relation
corresponding to the 3D estimate, $j^{3D}_{\rm tot}$, is well below the
other relations, because it does not mix bulge stars within the
rotating disk, as happens in the projected estimates.


\section{Comparison of the disk specific angular momentum to observations}
\label{app:j_disk}

 We compare the specific angular momenta of our model disks to the
  observational data by \citet{posti2018jdisk}, based on the SPARC galaxy
  sample \citep{lelli2016HI}.  These galaxies have a well studied rotational
  velocity profile, estimated from the HI. The specific angular momentum
  profile is integrated out to large radii (around $\sim 5 \,R_{\rm disk}$ most
  of the velocity profiles are found to converge). Since the sample is composed
  of gas rich spirals, we estimate the HI fraction in both observations and
  model galaxies, and estimate the influence of different selections on the
  final $j_{\rm disk}$--$M_*$ median relation.
 
\begin{figure*}
    \includegraphics[trim=2cm 0.8cm 2.5cm 1.5cm, clip, width = 0.7\linewidth]{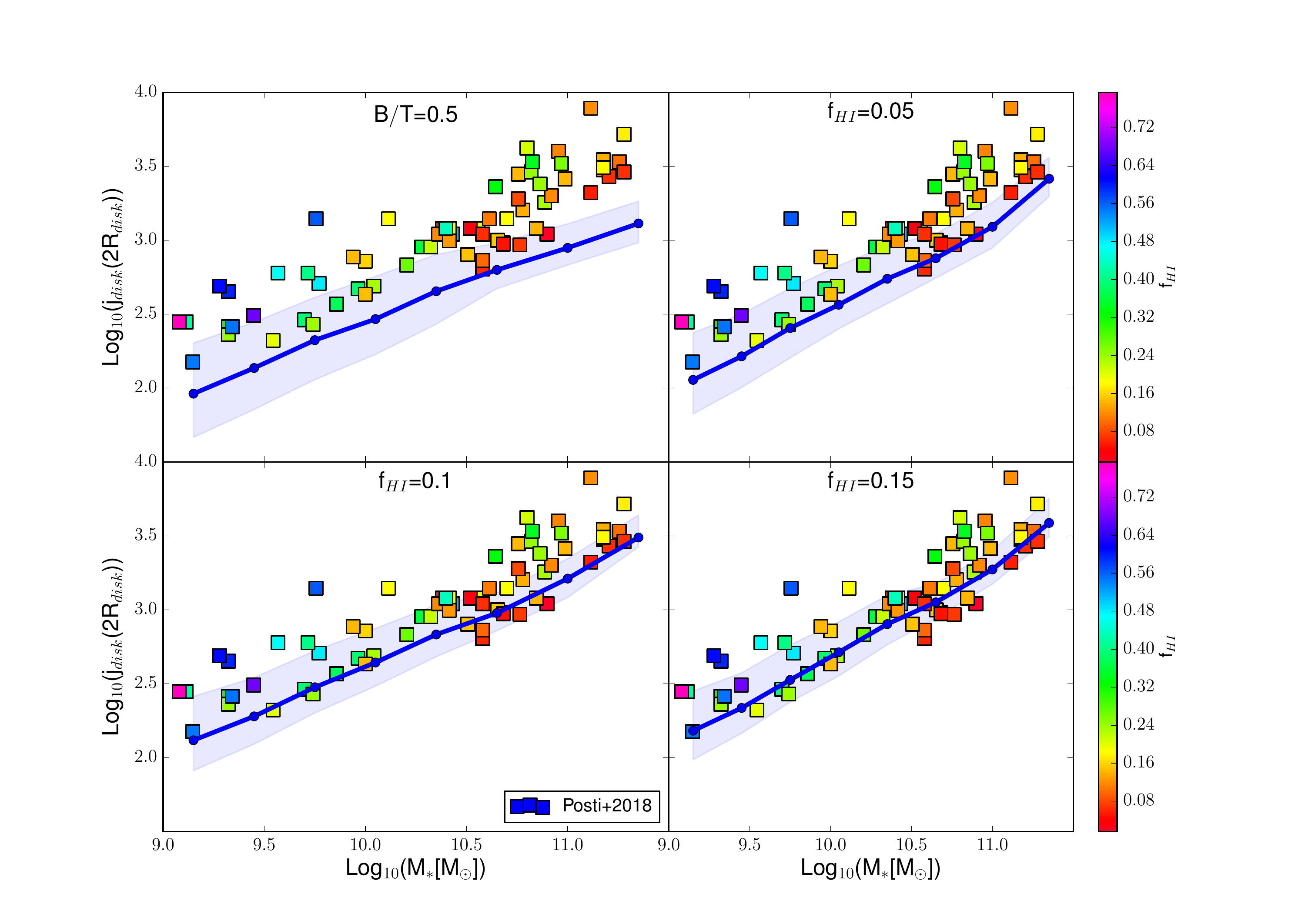}
    \caption{The $j_{\rm disk}$--$M_*$ relation predicted for LT galaxies (blue
      solid lines) from the X17MM run.  The $j_{\rm disk}$ value is integrated
      out to 2 disk scale radii.  Different panels show different LT
      selections.  Shaded areas show the region between the 16th and 84th
      percentiles of the distribution.  Symbols correspond to the observational
      measurements by \citet{posti2018jdisk}, colour-coded according to their
      $f_{\rm HI}$.}
    \label{fig:jdisk_ms}
\end{figure*}

Fig.~\ref{fig:jdisk_ms} shows the median $j_{\rm disk}$--$M_*$ relation (solid
blue line) we obtain for galaxies selected using $B/T<0.5$ (top left panel),
and an additional threshold constraint for the HI fraction $f_{\rm HI}=M_{\rm
  HI}/(M_*+M_{\rm HI})$ (other panels, as indicated in the captions). 
The model median relation corresponds to the
3D specific angular momentum of the stellar disks, evaluated out to 2 scale
radii.
Squares represent the observational data by \citet{posti2018jdisk}, color-coded
according to their $f_{\rm HI}$. 
Observational data show a clear correlation between HI gas fraction and $j_{\rm disk}$, with HI-rich galaxies having a higher specific angular momentum than gas-poor galaxies at fixed stellar mass. 
Furthermore, the HI gas fraction correlates with stellar mass, with higher stellar masses corresponding to  lower average HI fractions.
The model predicted median relation shifts upwards, particularly at the high
mass end, when imposing a threshold on the cold gas fraction. When using a cut
that is similar to that of the observational sample, model predictions agree relatively well with data. 

\begin{figure*}
    \includegraphics[trim=2cm 0.8cm 2.5cm 2cm, clip, width = 0.7\linewidth]{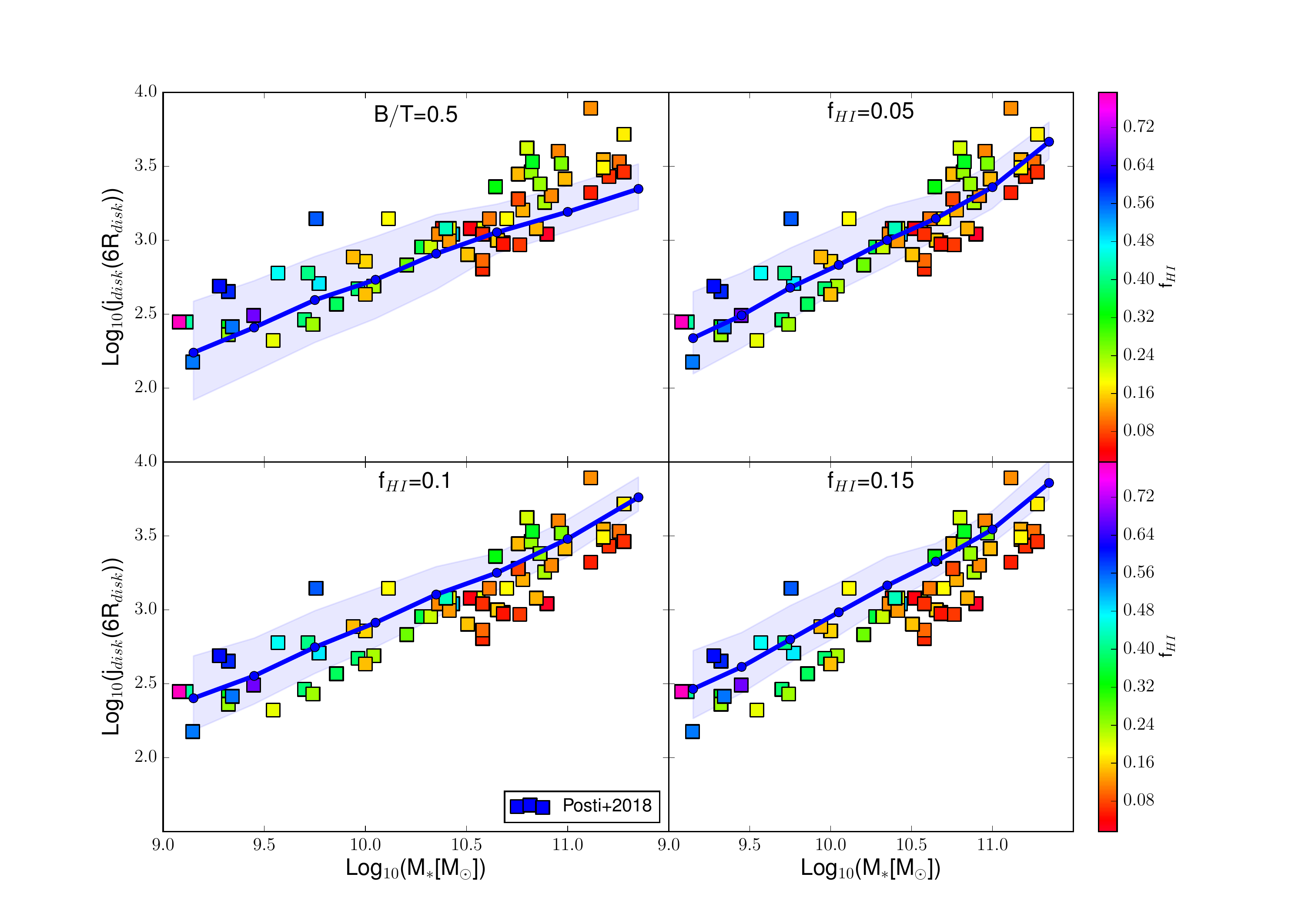}
    \caption{As in Fig.~\ref{fig:jdisk_ms}, but integrating the model specific
      angular momentum out to 6 disk scale radii.}
    \label{fig:jdisk_ms_r6}
\end{figure*}

We also integrated $j_{\rm disk}$ out to 6 disk scale radii, assuming this
length as sufficient for convergence. Results are shown in
figure~\ref{fig:jdisk_ms_r6}.  The median relation is shifted upwards by 0.2
dex for all the selections. This does not affect significantly our conclusions.

\section{Morphology and cold gas content}
\label{app:fcold}
\begin{figure*}
\centering
\includegraphics[trim=2.5cm 3cm 3cm 4cm, clip, width =
0.9\linewidth]{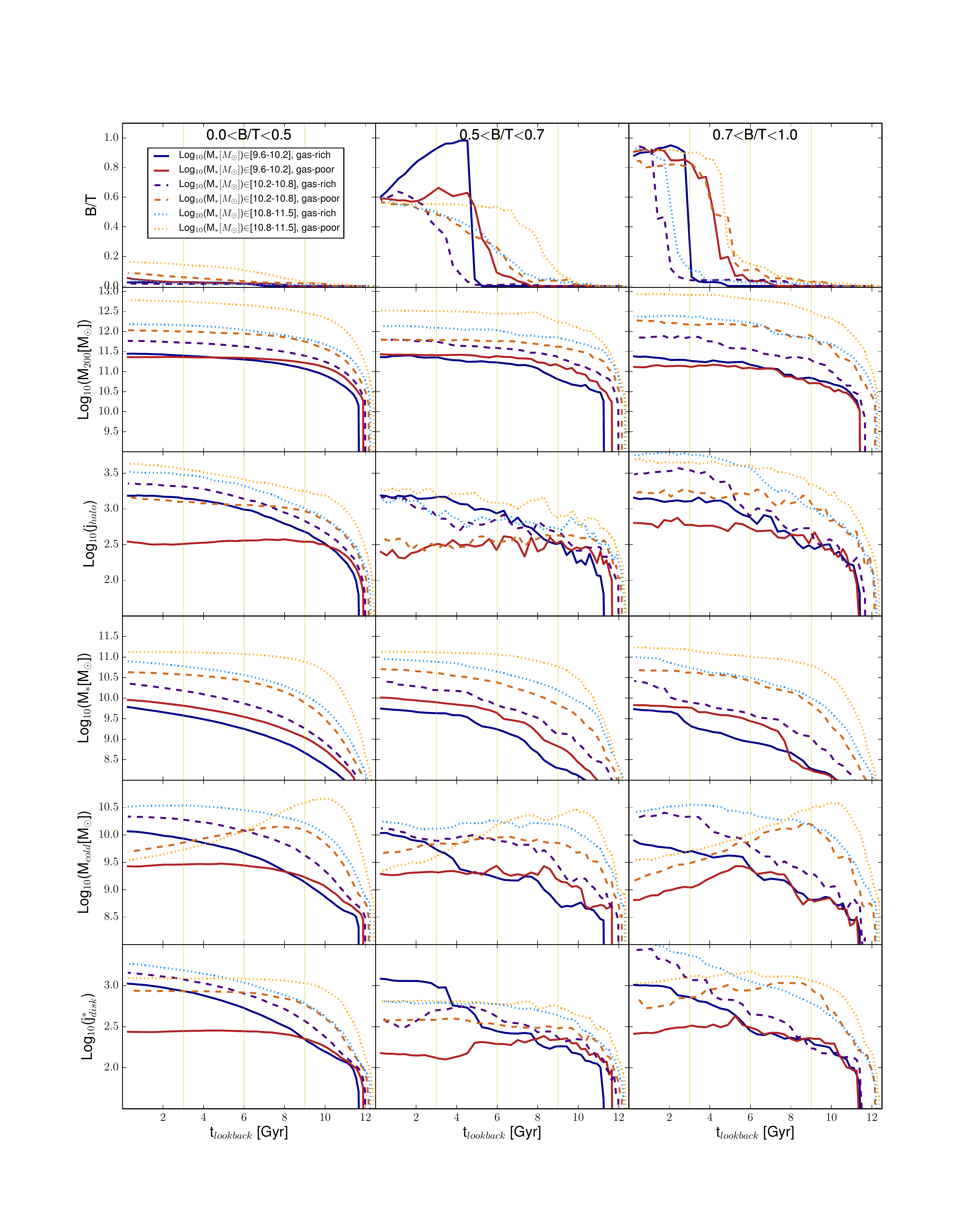}
\caption{Median evolution as a function of lookback time of several
  galactic properties.  From top to bottom: B/T, $M_{200}$,
  $j_{\rm halo}$, $M_*$, $M_{\rm cold}$, $SFR$, and stellar disk specific
  angular momentum.  Model galaxies have been selected according to
  their stellar mass at redshift 0, as indicated in the legend. A
  further binning is made as a function of the cold gas fraction
  $f_{\rm cold}$, with gas-rich (indigo, purple and pink) and gas-poor
  galaxies (chocolate, orange and gold) selected as those above or
  below the upper or lower 16th percentile of the distributions,
  respectively.  Different colors are used for galaxies of different
  stellar mass, as indicated in the legend.}
    \label{fig:history_BT_fgas_more}
\end{figure*}

Fig.~\ref{fig:history_BT_fgas_more} is similar to
Fig.~\ref{fig:history_fgas}, but considering a further binning of
model galaxies according to their $B/T$ (different columns).
In Sec.~\ref{sec:history_gas}, we have shown that the higher specific
angular momentum of gas-rich galaxies, with respect to gas poor ones,
is due to a larger contribution from recent star formation. This, in
turn, is due to the different accretion histories of their hosting
haloes. When considering galaxies divided in bins of $B/T$, we find an
evolution similar to that found for the entire population. 
The figure shows that gas-poor ET galaxies, at redshift zero, contain
less gas than LT galaxies. In addition, the bulges of gas-poor
galaxies form on average 2-3 Gyrs earlier than those of gas-rich
galaxies.  This is a selection effect, due to the fact that galaxies
with higher star formation rates can regrow a disk more efficiently.
As star formation correlates with the gas content, the star formation
rates are larger in gas rich galaxies, that can quickly regrow a
significant disk component.  A certain $B/T$ threshold selects
gas-poor ET galaxies that formed their bulge earlier than gas-rich
galaxies.  This is because the latter, in the meanwhile, have already
regrown their disks.

\bsp	
\label{lastpage}
\end{document}